\documentclass[aps,reprint,10pt,superscriptaddress,showpacs]{revtex4-1}
\usepackage[T1]{fontenc}
\usepackage[utf8]{inputenc}
\usepackage{color}
\usepackage{amsmath}
\usepackage{amssymb}
\usepackage{graphicx}
\usepackage[colorlinks, linkcolor=blue,anchorcolor=blue,citecolor=blue,urlcolor=blue]{hyperref}

\usepackage{dcolumn}
\usepackage{multirow}
\usepackage{bm}
\usepackage[none]{hyphenat} 
\usepackage{bbm}
\usepackage{braket}
\usepackage{tabularx}

\input{epsf}

\begin{document}

\title{Second-order and real Chern topological insulator in twisted bilayer $\alpha$-graphyne } 
\author{Bin-Bin Liu}
\affiliation{School of Physics, Beihang University, Beijing 100191, China}

\author{Xu-Tao Zeng}
\affiliation{School of Physics, Beihang University, Beijing 100191, China}

\author{Cong Chen}
\affiliation{School of Physics, Beihang University, Beijing 100191, China}
\address{Department of Physics, The University of Hong Kong, Hong Kong, China}

\author{Ziyu Chen}
\email{chenzy@buaa.edu.cn}
\affiliation{School of Physics, Beihang University, Beijing 100191, China}

\author{Xian-Lei Sheng}
\email{xlsheng@buaa.edu.cn}
\affiliation{School of Physics, Beihang University, Beijing 100191, China}
\affiliation{Peng Huanwu Collaborative Center for Research and Education, Beihang University, Beijing 100191, China}

\begin{abstract}
The study of higher-order and real topological states as well as the material realization have become a research forefront of topological condensed matter physics in recent years. Twisted bilayer graphene (tbG) is proved to have higher-order and real  topology. However whether this conclusion can be extended to other two-dimensional twisted bilayer carbon materials and the mechanism behind it lack explorations. In this paper, we identify the twisted bilayer $\alpha$-graphyne (tbGPY) at large twisting angle as a real Chern insulator (also known as Stiefel-Whitney insulator) and a second-order topological insulator. 
Our first-principles calculations suggest that the tbGPY at 21.78$^\circ$ is stable at 100 K with a larger bulk gap than the tbG.
The non-trivial topological indicators, including the real Chern number and a fractional charge, and the localized in-gap corner states are demonstrated from first-principles and tight-binding calculations. 
Moreover, with $\mathcal C_{6z}$ symmetry, we prove the equivalence between the two indicators, and explain the existence of the corner states. 
To decipher the real and higher-order topology inherited from the Moir\'e heterostructure, we construct an effective four band tight-binding model capturing the topology and dispersion of the tbGPY at large twisting angle. 
A topological phase transition to a trivial insulator is demonstrated by breaking the $\mathcal C_{2y}$ symmetry of the effective model, which gives insights on the trivialization of the tbGPY as reducing the twisting angle to 9.43$^\circ$ suggested by our first-principles calculations.
 
\end{abstract}

\maketitle
\section{INTRODUCTION}
The discovery of the topological insulator (TI) has stimulated a vibrant research field in condensed matter physics~\cite{Hasan_RMP,Qi_RMP,ShunQingShen_TI,Bansil_RMP}. A TI in $d$ dimensions has an insulating bulk, but features topologically protected gapless states on its $(d-1)$-dimensional (D) boundaries. Recently, the concept was extended to a novel class of topological phase-the higher-order TI~\cite{ZhangFan_PRL2013,Hughes2017,Langbehn2017,SongZD2017,Hughes2017b,Schindler2018SA}. An $n$th-order TI has topological gapless states at its $(d-n)$D  boundaries, but is gapped otherwise. For instance, a second-order topological insulator (SOTI) in 2D hosts topological gapless states at its 0D corners between its edges that are gapped. 
Previous works first reveal the higher-order TIs in 3D  materials~\cite{Bradlyn2017,Schindler2018,Schindler2018SA,Yue2019ws,WangZhijun_arXiv,Zhang2019tp,Vergniory2019ub,TangNP,TangNature,XuYF2019}.  
Then in 2D 
there are a few material candidates being proposed, such as the graphdiyne family~\cite{Sheng2019,WangZF2019,Lee2020th,chen2020GPY}, twisted bilayer graphene (tbG)~\cite{Park2019,Liu2021}, black phosphorene~\cite{Ezawa2018P,Hitomi2021}, Bi/EuO~\cite{Chen2020}, monolayer group-V~\cite{Radha2020,Huang2021}, and group-IV materials~\cite{Qian2021,Pan2022}. It is still a big challenge to find more realistic 2D SOTI materials.

With the spacetime inversion symmetry, the wavefunction over the Brillouin zone (BZ) is real instead of complex. The so-called real topology is characterized by the real Berry bundles over the BZ and can be indicated by the Real Chern number (RCN)\cite{RCN2017} or the second Stiefel-Whitney number~\cite{Nakahara2003}. 
Twisting one layer of the AA-stacking periodic lattice structures with translational symmetry, one can obtain a larger super cell which forms the so-called Moir\'e pattern. In the field of twistronics, the tbG is a pioneer material that has been investigated in many aspects including topology. 
The tbG Moir\'e system is shown to be topological~\cite{Song2019} and features a nontrivial RCN or the second Stiefel-Whitney number~\cite{Ahn2019,Park2019}. 
An insulator with nontrivial RCN (or second  Stiefel-Whitney number) like tbG can be referred  as a real Chern insulator (RCI)~\cite{Zhao2020} (or Stiefel-Whitney insulator~\cite{Lee2020th}).
Besides, the tbG system at relatively large Moir\'e twisting angle, say 21.78$^\circ$, is further proved to be a SOTI by explicitly demonstrating the existence of the corner states~\cite{Park2019}. 

By inserting the acetylenic linkage into graphene lattice, many carbon allotropes can be constructed
\cite{Baughman1987,GPY2012}. 
Except for graphene, other 2D carbon allotropes with twisted bilayer structures are lacking of  exploration, such as ($\alpha,~\beta,~\gamma$)-graphyne~\cite{GPY2012} and graphdiyne~\cite{LiYL2010,GDYreview2017,LiYL2017,ZhangJ2019}. These materials with hexagonal lattices all have the same space group symmetry as graphene, but their properties are not identical to graphene and vary in their own way due to the presence of the acetylenic linkage\cite{Jana2019}. So their twisted bilayer structures do not necessarily behave like the tbG. 

\begin{figure}
  \includegraphics[width=8.6cm]{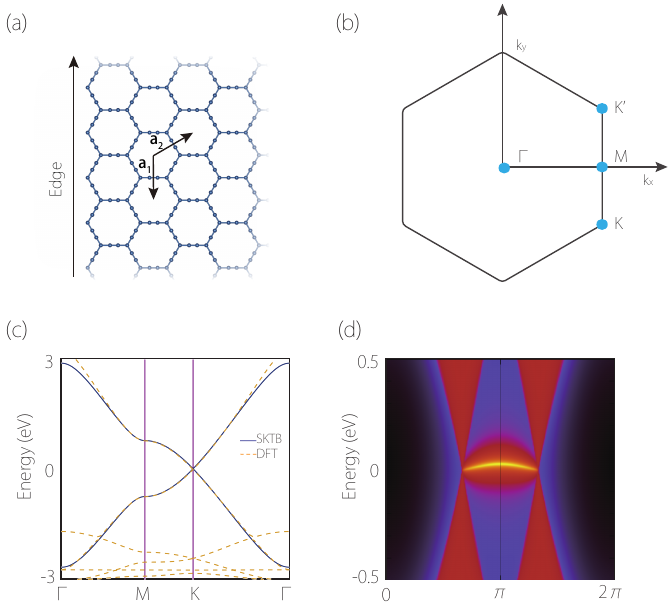}
  \caption{(a) Crystal structure and unit cell of the $\alpha$-graphyne with a zigzag edge. (b) shows the  high symmetry points in the first Brillouin zone used in our calculations. (c) and (d) show the bulk and edge band structures for the monolayer $\alpha$-graphyne. }
\label{Fig_monolayer}
\end{figure}

Among these graphynes, the $\alpha$-graphyne is the most similar one to graphene.
As studied in Refs.~\cite{GPY2012,Kim2012,Band2020,Jana2021}, $\alpha$-graphyne has eight carbon atoms per unit cell [see Fig.~\ref{Fig_monolayer}(a)], which can be regarded as inserting an acetylenic linkage between every two atoms in graphene. 
The symmetry group for $\alpha$-graphyne (and graphene) is the No. 191 space group with point group D$_{6h}$ ($p$6m symmetry). The $\alpha$-graphyne shares some properties with graphene. 
For example, the band structure of the $\alpha$-graphyne features a similar linear gapless cone at the K in the BZ as shown in  Fig.~\ref{Fig_monolayer}(c). 
Therefore, it is natural to ask if the twisted bilayer $\alpha$-graphyne (tbGPY for short) can inherit the second-order and real topology from  the Moir\'e heterostructure like the tbG at large twisting angle.

In this paper, we address the issue by exploring the electronic and topological properties of the tbGPY. Based on tight-binding and first-principles calculations, we identified the tbGPY at large twisting angle as a RCI and a SOTI, which has a larger bulk gap than the tbG. 
We confirmed the bulk topology from both the nontrivial RCN $\nu_{R}$~\cite{RCN2017,chen2020GPY} and a fractional charge $Q^{(2)}_{corner}$~\cite{Benalcazar2019} that indicates a filling anomaly. 
With the $\mathcal C_{6z}$ symmetry, we strictly proved that the nontrivial RCN equals an e/2 fractional  
charge $Q^{(2)}_{corner}$, which in turn suggests the existence of corner states on a hexagonal lattice. 
We further calculated the energy level of a hexagonal nano-disk for the tbGPY at 21.78$^{\circ}$ and found six in-gap localized corner states, which exemplifies the bulk-boundary correspondence of a RCI with $\mathcal C_{6z}$ symmetry. The robustness of the corner states against symmetry breaking disorders was demonstrated. 
We also performed an \textit{ab initio} molecular dynamics calculation which suggests that the tbGPY at 21.78$^{\circ}$ is stable at 100 K.

Moreover, we built an effective tight-binding model to decipher the SOTI and RCI state inherited from the Moir\'e heterostructure of tbGPY. The model features a RCI as well as a SOTI with non-trivial fractional charge, and can be transited to a trivial insulator by breaking $\mathcal C_{2y}$ symmetry. 
To extend, the model can be applied to other twisted bilayer materials with a Dirac cone at the K point in the monolayer, such as the tbG. 
Different from the tbG, we observed that the RCI state in tbGPY becomes trivial as the twisting angle becomes smaller. This trivialization can be simulated by our effective model through the topological phase transition.

\section{Lattice and band structures}
We begin by introducing the general geometry for the twisted bilayer Moir\'e system with hexagonal lattice. To obtain the Moir\'e pattern, one may stack two periodic lattice planes together (i.e. AA-stacking), and then twist one layer with respect to the other with respect of certain commensurate angle. For the hexagonal lattice unit cell, the commensurate angle can be formulated as   $\theta_i=\arccos(\frac{3i^2+3i+0.5}{3i^2+3i+1})$ (where $i$ is an integer above zero)~\cite{PhysRevLett.99.256802}. 
The commensurate condition does not depend on the inversion center, and is chosen as the middle of the hexagon lattice [Fig. \ref{Fig_monolayer} (a) ] in this work. The Moir\'e pattern and the associated lattice vectors only depend on the twisting angle or the index $i$.
In the following, we  use the $i$ to label the commensurate angle.
\begin{figure}
  \includegraphics[width=8.6cm]{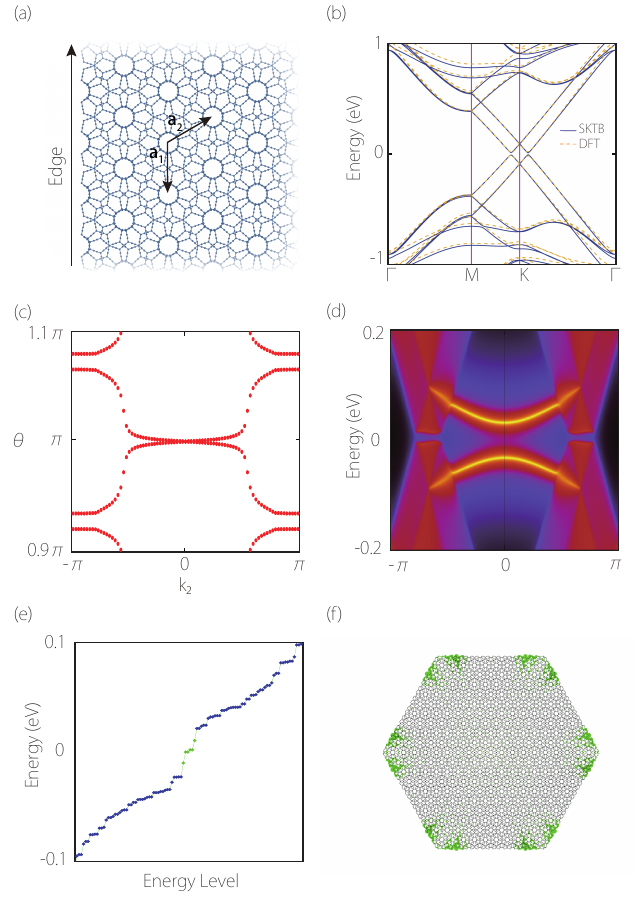}
  \caption{The  tbGPY ($i=1$) material and its properties. (a) Crystal structure with lattice vectors and a specified armchair edge. (b) and (d) show the gapped bulk and edge spectra with a gap around 13 meV for the bulk and  65 meV for the edge. (c) The Wilson loop spectrum around $\theta=\pi$, which indicats the non-trivial RCN ($\nu_{R}=1$). (e) The energy levels of the hexagonal nano-disk in (f) near the fermi-level.  (f) The real space distribution of the six in-gap states marked as green dots in (e), suggesting the localized corner states.}
\label{Fig_twisted_bilayer}
\end{figure}

The lattice structure of the twisted bilayer $\alpha$-graphyne at the commensurate angle 21.78$^\circ$ with $i=1$ is shown in Fig.~\ref{Fig_twisted_bilayer}(a), which is the main focus of the paper. 
To test the dynamical stability, we performed a molecular dynamics simulation for the tbGPY and find that it is dynamically stable under at least 100 K. The simulation result is summarized in  Appendix \ref{MD}.
Its Moir\'e supercell contains 112 carbon atoms, seven times as large as the AA-stacking $\alpha$-graphyne. 
We obtained the layer distance of the tbGPY with a first-principles relaxation calculation, which is $d_{0}=$3.444 \AA, in agreement with previous studies\cite{Leenaerts2013}. See Appendix \ref{MethodsFP} for calculation details. 
Different from the monolayer $\alpha$-graphyne, the tbGPY lies in the No. 177 space group with D$_6$ point group symmetry. The generators for this associated point group can be chosen as $\mathcal C_{6z}$ and $\mathcal C_{2y}$. 

We calculated the energy dispersion of the tbGPY with $i=1$ based on density functional theory. The result is shown in the dashed lines in Fig.~\ref{Fig_twisted_bilayer}(b), which is calculated along the line connecting the high symmetry points specified in Fig. \ref{Fig_monolayer} (b). We found a sizable gap of around 12.8 meV. To compare, we also calculate the energy spectrum for the twisted bilayer graphene with layer distance 3.35 \AA~using the same calculation method, and find a gap of around 1.4 meV at K. See Appendix \ref{MethodsFP} for explanations of the method. 
The opening of the band gap for the tbGPY can be understood from the $U_v(1)$ valley symmetry breaking for the bulk cones.

To proceed, we constructed a symmetry-based tight binding model with the method introduced by Slater and Koster\cite{Slater1954}. The details for the method are shown in Appendix~\ref{MethodsSK}. Following the method, 
we constructed a Slater Koster tight-binding (SKTB) model for tbGPY with one $p_z$ orbital per site. 
The fitted energy dispersion for the tbGPY using the SKTB model is shown in Fig.~\ref{Fig_twisted_bilayer}(b) with solid lines, which captures the main feature of the energy dispersion of tbGPY in the low energy range. 

The parameters for the SKTB model are illustrated as follows. The SKTB model includes all the couplings that have bond lengths less than the range of $L=5$ \AA. The nearest neighbor distance in the monolayer $\alpha-$graphyne is $a_0=1.232$ \AA,  and the layer distance of the tbGPY is $d_{0}=$3.444 \AA.
The nearest neighbor \textit{p}-\textit{p} $\pi$ coupling is fitted as $V_{pp\pi}^{0}=-4.45$ eV while the nearest \textit{p}-\textit{p} $\sigma$ coupling is $V_{pp\pi}^{0}=0.32$ eV. 
Other orbital couplings are determined relatively to those nearest ones by their relative distance to $a_0$ and $d_0$ [See Eq. (\ref{hopping_pi}) and Eq. (\ref{hopping_s})]. 
The decay length of the transfer integral is chosen as $\delta_0$=0.32 \AA. The onsite energy for each orbitals is fitted as -0.156 eV. 
In the following, we investigate the topological properties of the material.

\section{Real Chern number}
For a spinless system, under the spacetime inversion symmetry $\mathcal P\mathcal T$, the wavefunction over the BZ is real instead of complex. In such case, the topology is characterized by the real Berry bundles over the BZ and can be termed as the real topology. The topological indicator for the real topology is the the RCN\cite{RCN2017}, or the second Stiefel-Whitney number~\cite{Nakahara2003}.
The RCN is the defining topological invariant for the RCI as well as other exotic topological phase such as the second-order nodal-line semimetal\cite{Zhao2020,Chen2022}.  In 2D the defining spacetime inversion symmetry $\mathcal P\mathcal T$ can be replaced with the $\mathcal C_{2z}\mathcal T$ symmetry. Originated from the $D_{6}$ point group, the tbGPY has the  $\mathcal C_{2z}$ symmetry, while also possessing the time reversal symmetry $\mathcal T$. So it has the combined symmetry $\mathcal C_{2z}\mathcal T$ required for the real topological classification. 

In practice, there are two ways to compute the RCN~\cite{chen2020GPY}. One is the well-known Wilson loop method: calculating the Wilson loop along a chosen direction e.g., $\mathbf k_1$, with fixed the $k_2\in[-\pi,\pi)$. 
This results in N (the number of valence bands of the system) curves in the $k_2-\theta$ diagram, which represents the evolution of the Wannier center. The parity of the winding number or equivalently the RCN ($\nu_R$) can be read off from the times of crossings ($\zeta$) of the Wilson loop spectrum with the $\theta=\pi$ line\cite{Ahn2018}, i.e., 
\begin{eqnarray}\label{RCN_Wilson}
\nu_R=\zeta ~\text{mod~2}.
\end{eqnarray}
When not only the $\mathcal{PT}$ but also the $\mathcal P$ symmetry is preserved, there is another more intuitive method to calculate the RCN, which is by counting the parity eigenvalues of the valence bands at the four time reversal invariant momenta (TRIM) points $\Gamma_i$. Practically, one may follow
\begin{eqnarray}\label{RCN}
(-1)^{\nu_R}=\prod_{i=1}^4 (-1)^{\lfloor(n_-^{\Gamma_i}/2)\rfloor},
\end{eqnarray}
to compute the RCN $\nu_R$, where the $n_-^{\Gamma_i}$ is the number of the minus parity in the valence band at $\Gamma_i$.

We calculated the RCN for tbGPY with both methods. With the fist method, we obtained the Wilson loop from the SKTB model of tbGPY as in Fig.~\ref{Fig_twisted_bilayer}(c), where a single crossing of the spectrum with $\pi$ suggests the RCN to be nontrivial, i.e., $\nu_R=1$. 
With the second parity counting method, we calculate the parity eigenvalues of the $\mathcal C_{2z}$ at TRIM points from first-principles calculations. Using Eq. (\ref{RCN}), the RCN is also found to be nontrivial. 
The parity counting method can be further simplified in our case. As the tbGPY lattice possesses the $\mathcal C_{6z}$ symmetry, three of the TRIM points (M points) are equivalent, so that we only need to calculate one M point instead all three of them. Detailed information about the RCN calculation is shown in Tab.~\ref{RCN_tbGPY} in Appendix \ref{Appen_RCN}. Our results suggests that the tbGPY system is a RCI at large commensurate angle.

The second parity counting method is more physically intuitive, because it tells us information about band inversion. For example, a nontrivial RCN in Tab.~\ref{RCN_tbGPY} corresponds to a double band inversion at the $\Gamma$ point.

A nontrivial RCN in a $\mathcal P\mathcal T$ invariant system could feature a rich boundary correspondence\cite{Zhao2020}. This means that despite having a nontrivial RCN, the tbGPY is not necessarily a SOTI.
In the following, we explore the bulk-boundary correspondence for the tbGPY system and prove that it is a SOTI with nontrivial fractional charge that features localized corner states in a hexagonal lattice.

\section{Bulk-boundary correspondence}

In 2D, a TI has topologically protected gapless edge states, while for a SOTI the edge state is gapped but the conducting corner charges in the edge gap emerge.
To proceed, we firstly calculate the edge spectrum for the tbGPY ($i=1$) and compare it with that of monolayer graphyne. In Fig.~\ref{Fig_twisted_bilayer}(d) and Fig.~\ref{Fig_monolayer}(d), we show the edge spectra for the tbGPY and the monolayer $\alpha$-graphyne with the edges illustrated in Fig.~\ref{Fig_twisted_bilayer}(a) and Fig.~\ref{Fig_monolayer}(a). 
Comparing the two, we find that instead of trivially stacking, the two bulk Dirac cones in tbGPY are shifted to higher and lower energies with a gap in between. 
Correspondingly, the edge states (bright yellow curves) that are associated with the bulk Dirac cones are also shifted upward and downward, resulting in an edge gap. 
The edge gap is found to be around 65 meV which is larger than the bulk gap of 13 meV [Fig.~\ref{Fig_twisted_bilayer}(d)].

Then we explore the property of the second-order boundaries (corners) of tbGPY on a hexagonal nano-disk geometry related by $\mathcal C_{6z}$ rotation symmetry. Inside the edge gap, we find six localized corner states in the nano-disk.
Specifically, we calculate the energy level of a hexagonal nano-disk of tbGPY with 33360 carbon atoms ($p_z$ orbitals), and find six states inside the edge gap marked as green dots, as shown in Fig.~\ref{Fig_twisted_bilayer}(e). 
The observed non-degeneracy of the energy levels of the six corner states is due to the slight breaking of chiral symmetry in the tbGPY and the size effect of the nano-disk. This is proved in our effective model analysis.
We plot the charge distribution of the six in-gap states and find that they are localized at six corners of the hexagonal nano-disk in Fig.~\ref{Fig_twisted_bilayer}(f). 

We now briefly introduce the concept of fractional charge to explain the presence of the six corner states.
The fractional charge is a bulk property that indicates the filling anomaly of electrons and is defined under crystalline symmetries. 
Under the $\mathcal C_{2z}$, a fractional charge termed $Q^{(2)}_{corner}$ can be defined\cite{Benalcazar2019} (see Appendix \ref{Disorder} for detail). 
Further considering $\mathcal C_{6z}$ symmetry as in the tbGPY case, we can relate the $Q^{(2)}_{corner}$ with the RCN in a neat form 
\begin{align}
Q^{(2)}_{corner}=e\frac{\nu_{R}}{2}.
\label{fractional charge2}
\end{align}
According to Ref. \cite{Benalcazar2019}, the fractional charge is a secondary indicator which is well defined with the vanishing of the polarization indicator $\bf P^{(2)}$. Remarkably, under the protection of $\mathcal C_{6z}$ symmetry, we find 
\begin{align}
\bf P^{(2)}=\bf 0.
\label{polarization2}
\end{align} 
This is illuminating since it indicates a correspondence between the RCI and a SOTI with localized corner states. 
A nontrivial RCN is a bulk property, which in the presence of $\mathcal C_{6z}$ is equivalent to a nontrivial well-defined fractional charge $Q^{(2)}_{corner}$ that corresponds to secondary boundary states related by $\mathcal C_{2z}$. 
Here we derive the correspondence (Eq. \ref{fractional charge2} and Eq. \ref{polarization2}) with $\mathcal C_{6z}$, therefore there can exist six localized corner states in general.
We further demonstrate that the tbGPY ($i=1$) owns nontrivial fractional charges of $Q^{(2)}_{corner}=e/2$ through first-principles. 
Therefore, the hexagonal nano-disk of tbGPY hosts six corner states, as shown in Fig. \ref{Fig_twisted_bilayer} (f) and in Fig. \ref{Fig_4BTB_model} (f) in  Sec. \ref{BBC_4BTB}.

The six corner charges are robust against disorders along the edge, provided that the bulk and edge energy gaps remain open. This is because that the disorder might shift the in-gap corner charges by only an integer, but leaves the fractional part intact. The detailed calculations for the tbGPY nano-disk with disorder can be found in Fig. \ref{Fig_disorder} in Appendix \ref{Disorder}.

We have demonstrated the bulk-boundary correspondence of the tbGPY ($i=1$) by showing that it is not only a RCI, but also a SOTI with localized in-gap corner states.

\section{Effective four-band model}
To decipher the higher-order and real topology inherited from the complex Moir\'e supercell of tbGPY with numerous atoms, we provide a simpler effective model with only four orbitals that captures the non-trivial topological properties.

\subsection{Model construction}\label{Model construction}
We build the effective TB model starting from the symmetries of tbGPY. By effective, we mean that the effective model captures the main features of the bulk and edge dispersions, possesses the nontrivial topology, i.e., the nontrivial RCN and fractional charge, and characterizes the localized corner states. 

\begin{figure}[t!]
\includegraphics[width=8.6 cm]{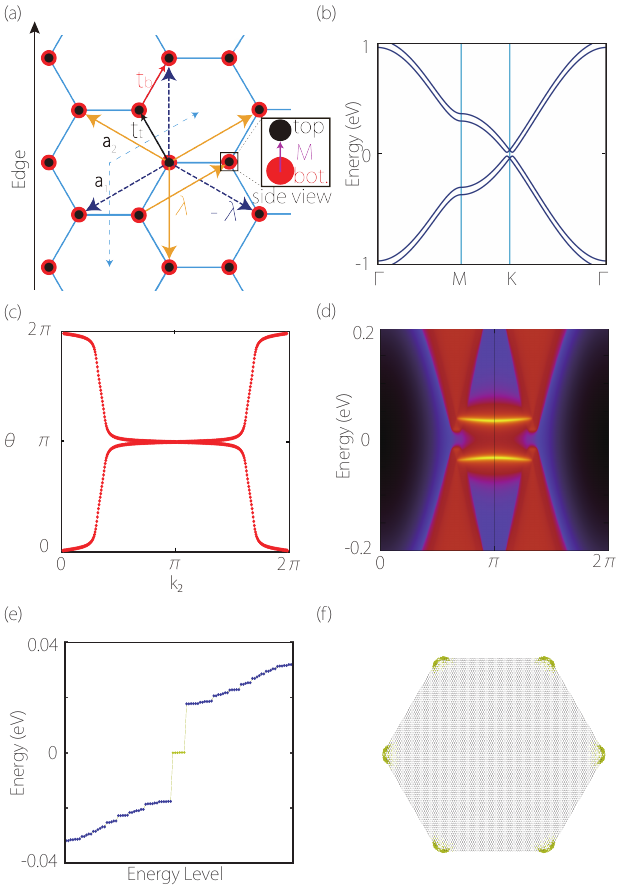}
\caption{ The topological 4BTB model and its properties. (a) Lattice, orbitals and specified hoppings with a zigzag edge. (b) and (d) show the gapped bulk and edge band structures. (c) The Wilson loop spectrum indicating the nontrivial RCN, $\nu_{R}=1$. (e) The energy levels of the hexagonal nano-disk in (f). (f) The charge distribution of the six in-gap states (marked as green dots) in
(e), indicating the presence of the localized corner states. The parameters are chosen as $t_t=t_b=t=1/3$ eV, $\lambda=0.01t$ and $M=0.1t$ for calculation. }
\label{Fig_4BTB_model}
\end{figure}

To achieve the requirement, we find the symmetries for constructing the effective model can be $\mathcal C_{2z}\mathcal T$ and crystalline $\mathcal C_{6z}$ and $\mathcal C_{2y}$. 
Due to the presence of the combined $\mathcal C_{2z}\mathcal T$ that defines the real topology, there exists a Wannier obstruction for building a two band tight binding model~\cite{Ahn2019,Zou2018}. Following the strategy in Refs.~\cite{Zou2018,Song2019}, we build a four band tight binding (4BTB) model in a honeycomb lattice, as shown in Fig.~\ref{Fig_4BTB_model}(a). The four orbitals of our 4BTB model forms two layers with each layer two sublattices in a unit cell. Here, we do not distinguish the orbital type, i.e., whether they are $s$ or $p_z$ orbitals. Without loss of generality, one may assume four $s$ orbitals with each orbital per site. Incorporating the symmetries into the orbitals, we can immediately specify the hoppings for the model as shown in Fig.~\ref{Fig_4BTB_model}(a).

Now we explain the coupling bonds for the 4BTB model. The black and red arrows are the nearest neighbor hoppings for the top and bottom layers, which generates two degenerated copies of massless Dirac cones at K point as that of graphene. The amplitudes of the hoppings are denoted as $t_t$ (top) and $t_b$ (bottom). The pink bond (denoted as  $M$) connects each of the nearest top and bottom orbitals as shown in the side view in Fig.~\ref{Fig_4BTB_model}(a). This bond functions as the energy shifting term which shifts the two Dirac cones upward and downward, which guarantees a gap for the edge spectrum. 
To have a bulk band gap around K, we introduce the alternative hoppings $\lambda$ that connect the top and bottom layer orbitals with solid orange and dashed dark blue arrows. 
The alternating pattern of the $\lambda$ is that the sign of the bond changes between positive and negative when rotating 60$^{\circ}$. These terms effectively introduce a gap in the bulk, corresponding to the Moire twisting in tbGPY.
The 4BTB model in momentum space then reads
\begin{eqnarray}\label{4BTB_eq}
 H^{\text{4BTB}}\left(\mathbf{k}\right)=
 \begin{bmatrix}   
    G_{t} & M\sigma_{0}+iS\\
    M\sigma_{0}-iS & G_{b}
 \end{bmatrix}
\end{eqnarray}
with 
\begin{align}
G_{j} &=   t_{j}\sum_{i=1}^3 \left[\sigma_{x}\cos\left(\boldsymbol{\delta}_{i}\cdot\mathbf{k}\right)-\sigma_{y}\sin\left(\boldsymbol{\delta}_{i}\cdot\mathbf{k}\right)\right]~(j=t,b), \nonumber \\
S &= 2\lambda\sigma_{z}\sum_{i=1}^3\sin\left(\mathbf{d}_{i}\cdot\mathbf{k}\right).\nonumber
\end{align}
The $\sigma_0$ is a 2-by-2 identity matrix, and the $\sigma_{x,y,z}$ are Pauli matrices acting on the sublattice degrees of freedom.
The Hamiltonian in Eq. \ref{4BTB_eq} distinguishes top and bottom layers. 
The  $\boldsymbol{\delta}_1=\frac{1}{3}\mathbf{a}_1 +\frac{2}{3}\mathbf{a}_2$, $\boldsymbol{\delta}_2=-\frac{2}{3}\mathbf{a}_1-\frac{1}{3}\mathbf{a}_2$, and $\boldsymbol{\delta}_3=\frac{1}{3}\mathbf{a}_1 -\frac{1}{3}\mathbf{a}_2$ are the nearest hopping vectors within one layer, while $\mathbf{d}_1=\mathbf{a}_1$, $\mathbf{d}_2=\mathbf{a}_2$, and  $\mathbf{d}_3=-\mathbf{a}_1-\mathbf{a}_2$ are the second-nearest hopping vectors  within one layer. 
Note that the hopping parameters of the 4BTB model are all real numbers due to an emergent $\mathcal T$ symmetry. 
Therefore the 4BTB model recovers the same symmetries as tbGPY.

\subsection{Nontrivial bulk-boundary correspondence}\label{BBC_4BTB}
To examine the validity of the effective 4BTB model, we check the criterion mentioned at the beginning of the Sec.\ref{Model construction}. Specifically, we set $t_t=t_b=t=1/3$ eV, $\lambda=0.01t$ and $M=0.1t$ for the model. The parameters are obtained by simulating the main features of the bulk band of tbGPY around K. 

The bulk property concerns band dispersion as well as the topological indicators.
The band structure has a small gap around K with a valley splitting feature, in agreement with the material calculations of tbGPY, as shown in Fig.~\ref{Fig_4BTB_model}(b) and Fig.~\ref{Fig_twisted_bilayer}(b). 
The RCN as well as the fractional charges $Q^{(2)}_{corner}$ are all nontrivial and agree with the tbGPY ($i=1$) as shown in the $i=1$ row of Tab. \ref{RCN_Q}. 
 We also calculate the RCN with the Wilson loop technique. We calculate the Wannier center evolution along the $k_2$ which has two curves and a single crossing point at $\theta=\pi$, which means the RCN is nontrivial. See  Fig.~\ref{Fig_4BTB_model}(c).
Therefore, the topological 4BTB model features a RCI with nontrivial fractional charges.

\begin{table}
\caption{\label{RCN_Q}
The RCNs and fractional charges for tbGPY material and effective 4BTB model. The first two columns indicate the systems with different $i$. The remaining columns are the RCN $\nu_{R}$ and the fractional charge of $Q^{(2)}_{corner}$ 
calculated by Eq.~(\ref{RCN}) and (\ref{fractional charge2}). 
The first two rows are tbGPY with $i=1$ calculated from first-principles and simulated by the topological  4BTB model. The last two rows are the tbGPY with $i=3$ calculated from first-principles and simulated by the trivial 4BTB model (which differs from the topological 4BTB model in $t_{b} = -t_{t} =-1/3 $ eV).  }
\begin{ruledtabular}
\begin{tabular}{cccc}
$i$ & System & $\nu_{R}$  & $Q^{(2)}_{corner}$ \\ \hline
  \multirow{2}{*}{$i=1$} &   tbGPY     & 1  & e/2     \\
      & Topological  4BTB      & 1  & e/2    \\
\multirow{2}{*}{$i=3$} &    tbGPY     & 0  & 0     \\
      & Trivial 4BTB     & 0  & 0     \\
   \end{tabular}
\end{ruledtabular}
\end{table}

Then we check the boundary correspondence of the topological 4BTB model. 
We investigate the edge spectrum along the zigzag direction specified in Fig.~\ref{Fig_4BTB_model}(a), and the result is shown in Fig.~\ref{Fig_4BTB_model}(d). One observes that each edge connects two bulk cones, and leaves an edge gap in between, which has the same feature as that in Fig.~\ref{Fig_twisted_bilayer}(d). 
To investigate the second-order edge, we construct  a nano-disk containing 30000 orbitals for the 4BTB model. There are six degenerated zero modes at the Fermi level labeled by green dots, as shown in Fig.~\ref{Fig_4BTB_model}(e), which are distributed at the six corners of the nano-disk as shown in Fig.~\ref{Fig_4BTB_model}(f). In the model calculation, we are able to eliminate the size effect and hence the corner modes are nearly degenerated at 0 eV. 
See Appendix \ref{topological origin} for an explanation of topological origin of the corner state  with domain wall theory.

Therefore, we have constructed an effective model with only four orbitals for the tbGPY at large commensurate angle.  
To extend, the 4BTB model may also be applied to study other twisted bilayer system at large commensurate angle, such as tbG which has the same symmetry as tbGPY and has a Dirac cone at K in the monolayer~\cite{Park2019}.

\subsection{Topological phase transition}

After demonstrating the nontrivial topology of the 4BTB model, it is natural to explore the other side of the model, namely a trivial insulator.
We here provide a simple and intuitive way to  trivialize the 
effective 4BTB model. 
To achieve so, we simply reverse the sign of $t_{b}$, following that $t_{b} = -t_{t} = -t$. 
In this way, we break the $\mathcal C_{2y}$ symmetry of the 4BTB model and achieve a band inversion between the valence and conductance bands. 
Since the $\mathcal C_{6z}$ symmetry is still remained in the model, we can use Eq. \ref{RCN} and Eq. \ref{fractional charge2} to calculate the RCN and the fractional charge.
The topological indicators for the trivial 4BTB model are both 0 as shown in Tab.~\ref{RCN_Q}.
Slight breaking of the $\mathcal C_{2y}$ symmetry will not spoil the topological corner states as long as the bulk and edge gaps are not closed\cite{Sheng2019}. However, here the $\mathcal C_{2y}$ symmetry breaking is accompanied with a band inversion, which is the reason that the RCN and the fractional charge become trivial. As a consequence, there is no localized corner state in the trivial 4BTB model, in contrast to that of the topological 4BTB model as shown in Figs.~\ref{Fig_4BTB_model}(e, f). 

For tbGPY with smaller commensurate angel, say 9.4$^{\circ}$ with $i=3$, the RCN becomes trivial (see  Appendix \ref{Appen_RCN} for some details and  discussions). This indicates a band inversion. The trivial 4BTB model features the same RCN and fractional charge as the material with $i=3$ as presented in Tab.~\ref{RCN_Q}, and gives an understanding of the trivialization process from the symmetry breaking and band inversion prospects.

\section{Conclusion and outlook}
In conclusion, we proposed the tbGPY at large commensurate angle as a RCI and a SOTI from first-principles and tight-binding model calculations. The tbGPY has a sizable bulk gap of around 13 meV larger than the tbG and is stable at 100 K, which is favorable for the experimental detection of the corner states. 
To decipher the topological property inherited from the Moir\'e structure, we constructed a simple but  effective 4BTB model which captures the major topological features of the tbGPY.  The model may also apply to other twisted bilayer systems with a Dirac cone at K in the monolayer, such as the tbG. 
A phase transition of the effective model by breaking the $\mathcal C_{2y}$ symmetry is also studied, which gives insights on the trivialization of the RCI state in the tbGPY.  
Moreover, in the presence of $\mathcal C_{6z}$ symmetry, we prove that the nontrivial RCN $\nu_{R}$ equals an e/2 fractional charge $Q^{(2)}_{corner}$, which in turn explains the six localized corner states in the tbGPY hexagonal nano-disk.

We give an outlook on the experimental realization of the RCI and SOTI in artificial systems with the 4BTB model.  
In the topological 4BTB model, all the hopping amplitudes are real, either positive or negative, which preserves the time-reversal symmetry $\mathcal T$. Therefore, the model can be realized in a rich category of artificial systems, such as electrical circuit systems\cite{circuit2018}, phonon lattices\cite{Ma2019,Xue2020,Xue2019to}, photonic crystals\cite{Ozawa2019,Mittal2019}, mechanical systems\cite{mechanical2009}, and cold atoms\cite{ColdAtomRMP}. Moreover, a topological phase transition can be realized by simply reversing the sign of the nearest neighbor hoppings in the bottom/top layer $t_{b/t}$, which can be engineered to tune the existence of the corner states. 
In addition, removing the $M$ term in the 4BTB model, the remaining spinless model with pseudo-spin  generated by the layer operator $\bf\tau$ behaves effectively like the spinful model by Kane and Mele\cite{Kane2005Graphene}. 
Here the model can be realized with real hopping amplitudes and without spin degree which is favorable for realizations in artificial systems. 

\section*{Acknowledgments}
The authors thank Shengyuan A. Yang, Y. X. Zhao and Zhijun Wang for helpful discussions. This work is supported by the National Natural Science Foundation of China (Grants No. 12174018, No. 12074024, and No. 11774018).
\appendix
\renewcommand{\theequation}{A\arabic{equation}}
\setcounter{equation}{0}
\renewcommand{\thefigure}{A\arabic{figure}}
\setcounter{figure}{0}
\renewcommand{\thetable}{A\arabic{table}}
\setcounter{table}{0}
\section{Molecular dynamics study of the tbGPY}\label{MD}
\begin{figure}
  \includegraphics[width=8.6cm]{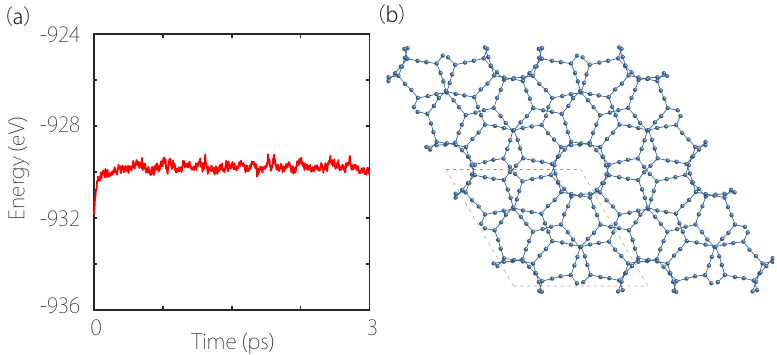}
  \caption{AIMD simulation results for tbGPY ($i=1$). (a) Variation of the total energy during the AIMD simulation at 100K. (b) Lattice structure at the end of the simulation period.}
\label{Fig_MD}
\end{figure}

To assess the dynamical stability of the tbGPY, we performed an \textit{ab initio} molecular dynamics (AIMD) simulation at 100K. With a time step of 1.5 fs, for total 3 ps. We find the variation of the total energy is small, and the structure of tbGPY and the Moir\'e pattern do not change too much after the simulation, as shown in Fig. \ref{Fig_MD}. Therefore, we can conclude that there are no obvious distortions of the geometries for tbGPY heterostructure at 100K, which suggests that it is dynamically stable at 100K.

\renewcommand{\theequation}{B\arabic{equation}}
\setcounter{equation}{0}
\renewcommand{\thefigure}{B\arabic{figure}}
\setcounter{figure}{0}
\renewcommand{\thetable}{B\arabic{table}}
\setcounter{table}{0}

\section{Methods}\label{Methods}
\subsection{First-principles calculations}\label{MethodsFP}
Our first-principles calculation is carried out based on the density-functional theory implemented in the Vienna \textit{ab initio} simulation package (VASP) \cite{Kresse1994,Kresse1996}. 
The projector augmented wave (PAW) method \cite{PAW} was used for treating the ionic potentials. The generalized gradient approximation (GGA) with the Perdew-Burke-Ernzerhof (PBE)~\cite{PBE} realization was adopted for the exchange-correlation functional. 
The plane-wave cutoff energy was set to 520 eV. Only the $\Gamma$ point is chosen for the BZ sampling in the self-consistent calculations, considering the large unit cell of the Moir\'e structure. 
A vacuum of around 20 \AA  is included in the simulation to reduce the unwanted interaction between any graphyne bilayers. 
The crystal structure optimization is stopped when the forces on the ions are less than 0.01 eV/\AA. The energy convergence criteria is set as 10$^{-6}$ eV for the electronic self-consistent calculations.
The van der Waals corrections are carried out in the relaxation calculations by the semiempirical density-functional theory (DFT)- D3 method\cite{DFTD3} with Becke-Johnson damping\cite{DFTD3BJ}.

To obtain the layer distance of tbGPY, we performed several full-atom structural relaxation with different initial interlayer distances ranging from 3.664 \AA  to 2.964 \AA  with an interval of 0.1 \AA. The free energy  after relaxation reaches a minimum of -938.3827 eV with the corresponding layer distance of 3.444 \AA, as presented in the main text.

\subsection{SKTB model}\label{MethodsSK}
We study the edge spectra and corner states using the SKTB model, following the Slater and Koster scheme~\cite{Slater1954,Moon2013}. 
We build the SKTB model with the $p_z$ orbital of the carbon atom in tbGPY. The general Hamiltonian of the SKTB model is written as
\begin{equation}
    H =-\sum_{|\mathbf r_{i}-\mathbf r_{j}|<L} h(\mathbf r_{i} - \mathbf r_{j})|\mathbf r_{i}\rangle\langle\mathbf r_{j}|+\text{H.c.}, 
\end{equation}
where $\mathbf r_{i}$ is the position of the lattice point i, and $|\mathbf r_{i}\rangle$ represents the atomic state at site i. The $L$ is the maximum length for the hopping considered. The $h(\mathbf r_{i} - \mathbf r_{j})$ is the hopping amplitude (transfer integral) between site i and j, which can be calculated using
\begin{align}
 -h(\mathbf d) =& V_{pp\pi}\left[1 - \left(\frac{\mathbf d \cdot \hat{\mathbf e}_z }{d} \right)^{2} \right]
+ V_{pp\sigma} \left(\frac{\mathbf d \cdot \hat{\mathbf e}_z}{d} \right)^{2},
\end{align} 
with
\begin{align}
V_{pp\pi} =& V_{pp\pi}^{0}\exp\left(-\frac{d-a_{0}}{\delta_{0}}\right) ,\label{hopping_pi}\\
V_{pp\sigma} =& V_{pp\sigma}^{0}\exp\left(-\frac{d-d_{0}}{\delta_{0}}\right)\label{hopping_s},
\end{align} 
where the $V_{pp\pi}^{0}$ is the nearest neighbor \textit{p}-\textit{p} $\pi$ coupling, and $V_{pp\sigma}^{0}$ is the nearest neighbor \textit{p}-\textit{p} $\sigma$ coupling. The nearest distance in the monolayer is $a_0$, and the layer distance is $d_{0}$. To consider the interlayer effect, $L>d_{0}$ should be satisfied. 

\renewcommand{\theequation}{C\arabic{equation}}
\setcounter{equation}{0}
\renewcommand{\thefigure}{C\arabic{figure}}
\setcounter{figure}{0}
\renewcommand{\thetable}{C\arabic{table}}
\setcounter{table}{0}
\section{Fractional charges and robustness of corner states against symmetry-breaking disorders}\label{Disorder}

The fractional charge quantization of corner states is due to the filling anomaly between the number of electrons required to satisfy the charge neutrality and the crystalline symmetry. 
The tbGPY preserves $\mathcal C_{2z}$ and $\mathcal C_{6z}$ symmetries which correspond to fractional charges of $Q^{(2)}_{corner}$ and $Q^{(6)}_{corner}$ defined as \cite{Benalcazar2019}
\begin{align}
Q^{(2)}_{corner} &= \frac{e}{4}\left[-n_{C_{2} }^{X} -n_{C_{2} }^{Y} +\left(n_{C_{2} }^{M} +n_{C_{2}}^{\Gamma }\right)\right] ~\text{mod}~e,\label{fractional charge20} \\
Q^{(6)}_{corner} &= e\left(\frac{ n_{C_{2} }^{M} -n_{C_{2}}^{\Gamma } }{4} + \frac{ n_{C_{3} }^{K} -n_{C_{3} }^{\Gamma }}{6}\right) ~\text{mod}~e
\label{fractional charge0},
\end{align} 
where the $n_{C_{k} }^{{i}}$ denotes the number of eigenvalue 1 of $\hat C_{k}$ (alone $\hat z$) at the point ${i}$ in the momentum space.
In the presence of $\mathcal C_{6z}$ symmetry, the $Q^{(2)}_{corner}$ admits a more intuitive form shown in Eq. \ref{fractional charge2} with Eq. \ref{polarization2}. As a result, the nontrivial RCN which is a bulk property corresponds to the fractional charge $Q^{(2)}_{corner}=e/2$ that features a unique boundary property.  
However, the $Q^{(6)}_{corner}$ does not give a one-to-one correspondence to the RCN, 
\begin{align}
Q^{(6)}_{corner} &= e\left(\frac{\text{sgn}\left(n_{C_{2} }^{M} -n_{C_{2}}^{\Gamma }\right)\nu_{R}}{2} + \frac{ n_{C_{3} }^{K} -n_{C_{3} }^{\Gamma }}{6}\right) ~\text{mod}~e
\label{fractional charge}.
\end{align} 
We find $Q^{(6)}_{corner}=e/6$ for the nontrivial tbGPY. 

\begin{figure}
  \includegraphics[width=4.3cm]{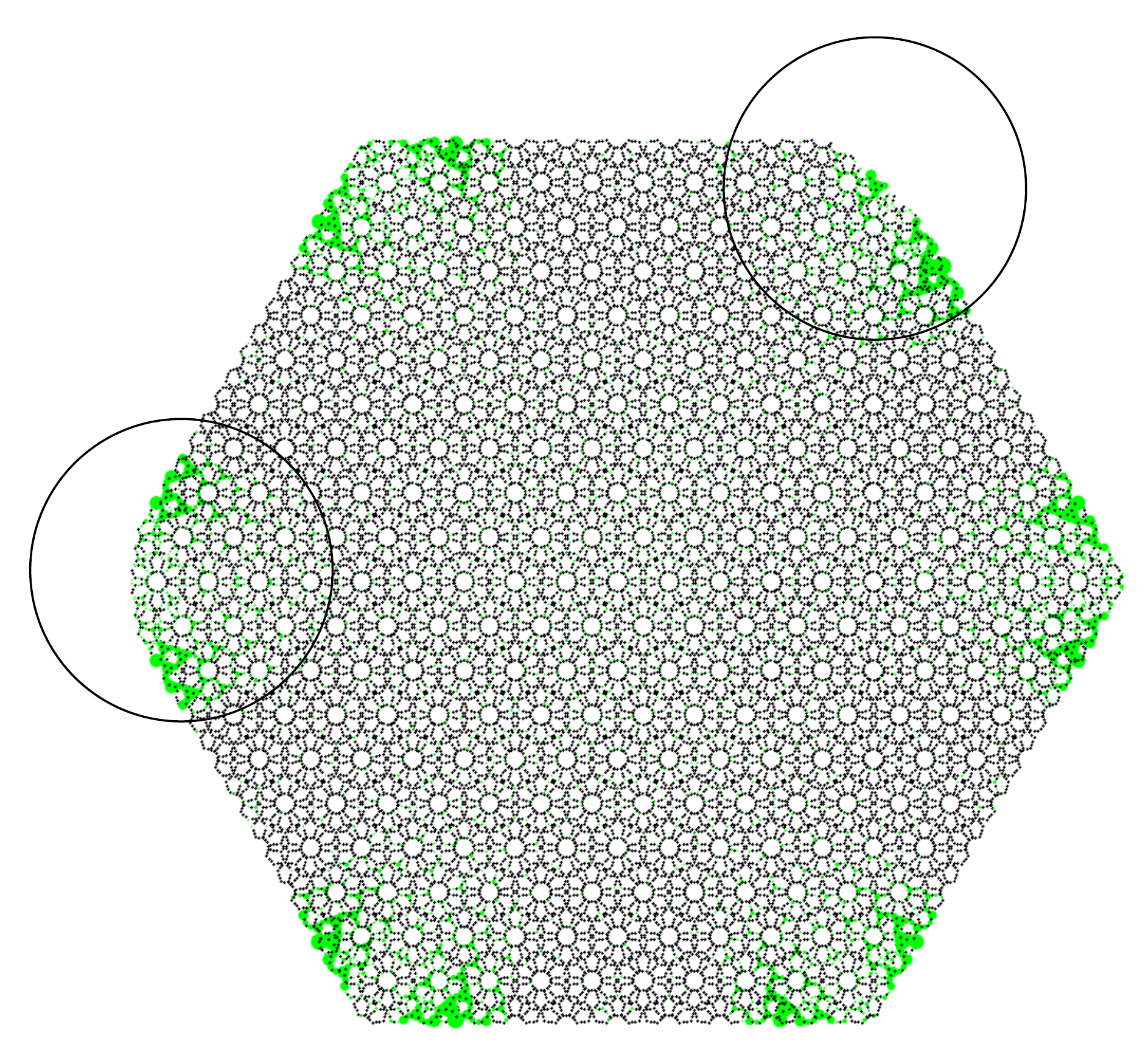}
  \caption{Robustness of corner states on a disordered hexagonal nano-disk of tbGPY $i=1$. The disorders are introduced at the corners enclosed by circles.}
\label{Fig_disorder}
\end{figure}

Then we explain this robustness from the point of view of fractional charge. 
In higher order topological insulator materials with disorders, if the crystalline symmetry, e.g., the $\mathcal C_{2z}$ or $\mathcal C_{6z}$, is slightly broken (which does not invert bands), it may cause a ground state filling of the in-gap states. This could shift the corner charges by only an integer, and therefore the fractional portion of the charge remains intact at each corners. 
We demonstrated the above analysis of the robustness of corner state by introducing disorders along the edge. Specifically, we removed several atoms at the corners (outlined in Fig. \ref{Fig_disorder}.
) of the hexagonal nano-disk. The disorder breaks the local $\mathcal C_{2z}$, $\mathcal C_{6z}$ and the $\mathcal C_{2y}$ crystalline symmetries in the nano-disk, but does not spoil the corner charges.

\renewcommand{\theequation}{D\arabic{equation}}
\setcounter{equation}{0}
\renewcommand{\thefigure}{D\arabic{figure}}
\setcounter{figure}{0}
\renewcommand{\thetable}{D\arabic{table}}
\setcounter{table}{0}

\section{RCN calculations for tbGPY and tbG with different twisting angles}\label{Appen_RCN}
\begin{table*}[!htbp]
\caption{\label{RCN_tbGPY}
The RCNs for tbGPY systems with i = 1–4 calculated from first principles. The first two columns are the commensurate twisting angles and their labels; the third column is the number of valence band for first-principles calculations. The forth and fifth columns are the number of minus parity at the $\Gamma$ and M counted from the valence bands. The last column suggests the RCNs calculated from the forth and fifth columns using Eq.~(\ref{RCN}). }
\begin{ruledtabular}
\begin{tabular}{cccccc}
i  & twisted angle ($^\circ$)&Number of valence band  &$n^\Gamma_-$  & $n^M_-$ 
&RCN\\ \hline
     1     & 21.787  & 224   & 110    & 112    & 1 \\
    2    & 13.174  & 608    & 302    & 304   & 1 \\
        3    & 9.430  & 1184    & 592    & 592    & 0 \\
            4     & 7.341  & 1952   & 976    & 976    & 0    \\
\end{tabular}
\end{ruledtabular}

\end{table*}

\begin{table*}[htbp]
\caption{\label{RCN_tbG}
The RCNs for tbG systems with several $i=$ 1-10 calculated from DFT. The arrangement of table is the same as Tab. \ref{RCN_tbGPY} }
\begin{ruledtabular}
\begin{tabular}{cccccc}
i  & twisted angle ($^\circ$)&Number of valence band  &$n^\Gamma_-$  & $n^M_-$ 
&RCN\\ \hline
    1     & 21.787  & 56    & 24     & 30     & 1 \\
    2     & 13.174  & 152    &72    & 78    & 1 \\
    3     & 9.430  & 296   & 144    & 150    & 1 \\
    4     & 7.341  & 488   & 240    & 246    & 1 \\
    5     & 6.009  & 728   & 360    & 366    & 1 \\
    6     & 5.086  & 1016   & 504   & 510   & 1 \\
    7     & 4.409  & 1352   & 672   & 678   & 1 \\
    8     & 3.890  & 1736   & 864   & 870   & 1 \\
    9     & 3.481  & 2168  & 1080   & 1086   & 1 \\
    10    & 3.150  & 2648  & 1320   & 1326   & 1 \\
\end{tabular}
\end{ruledtabular}
\end{table*}

Using Eq. \ref{RCN} in the main text with DFT, we  calculate the RCNs for tbGPY and tbG which are presented in Tab. \ref{RCN_tbGPY} and  \ref{RCN_tbG}.
We find a trivialization of RCN in the tbGPY with $i=3$ and 4. 
For the tbG systems, however, we have not observed such  trivialization of RCN up to $i=10$.

Now we discuss briefly the trivialization of the RCN in tbGPY with $i=3$. 
Despite the similarities between graphene and $\alpha$-graphyne, the existence of the acetylenic linkage makes  some differences in the $\alpha$-graphyne.  For instance, the acetylenic linkage in the $\alpha$-graphyne inverts the energy band, which in turn reverses the chirality between the graphene and the $\alpha$-graphyne at K (or K')\cite{Kim2012,Jana2019}. 
The trivialization in the RCN for tbGPY clearly indicates a double band inversion as reducing the twisting angle, while for graphene, such band inversion does not happen down to at least  3.15$^{\circ}$. This difference may be due to the existence of the acetylenic linkage in the $\alpha$-graphyne.

\renewcommand{\theequation}{E\arabic{equation}}
\setcounter{equation}{0}
\renewcommand{\thefigure}{E\arabic{figure}}
\setcounter{figure}{0}
\renewcommand{\thetable}{E\arabic{table}}
\setcounter{table}{0}

\section{Higher-order topological origin from domain wall theory}\label{topological origin}
The existence of corner states can also be understood in light of the domain wall theory. 
Previous studies~\cite{Moore2019,Sheng2019,Chen2020,chen2020GPY} suggest that a general SOTI system has the edge of the form
\begin{equation}
    H_{edge}({k}) = vk\sigma_z + m\sigma_x.
\end{equation}
If the mass term is odd under a reflection symmetry which connects two edges, then there should appear a domain wall state at the corner of the two edges. In the 4BTB model the $\pi$-rotation about the y axis represented as $\mathcal C_{2y}=\tau_x\sigma_{x}$ can serve as the reflection.  Under $\mathcal C_{2y}$, the velocity term reverses its sign with the mass term unchanged. Therefore, it suggests that the mass term is odd under the $\mathcal C_{2y}$, meaning that a domain wall state should appear at the corner where the two edges connected by the $\mathcal C_{2y}$ cross. Further considering the $\mathcal C_{3z}$ symmetry, there exist six corner states in the hexagonal nano-disk.



\bibliographystyle{apsrev4-1}{}
\bibliography{SOTI_TBGPY_ref}

\begin{thebibliography}{71}%
\makeatletter
\providecommand \@ifxundefined [1]{%
 \@ifx{#1\undefined}
}%
\providecommand \@ifnum [1]{%
 \ifnum #1\expandafter \@firstoftwo
 \else \expandafter \@secondoftwo
 \fi
}%
\providecommand \@ifx [1]{%
 \ifx #1\expandafter \@firstoftwo
 \else \expandafter \@secondoftwo
 \fi
}%
\providecommand \natexlab [1]{#1}%
\providecommand \enquote  [1]{``#1''}%
\providecommand \bibnamefont  [1]{#1}%
\providecommand \bibfnamefont [1]{#1}%
\providecommand \citenamefont [1]{#1}%
\providecommand \href@noop [0]{\@secondoftwo}%
\providecommand \href [0]{\begingroup \@sanitize@url \@href}%
\providecommand \@href[1]{\@@startlink{#1}\@@href}%
\providecommand \@@href[1]{\endgroup#1\@@endlink}%
\providecommand \@sanitize@url [0]{\catcode `\\12\catcode `\$12\catcode
  `\&12\catcode `\#12\catcode `\^12\catcode `\_12\catcode `\%12\relax}%
\providecommand \@@startlink[1]{}%
\providecommand \@@endlink[0]{}%
\providecommand \url  [0]{\begingroup\@sanitize@url \@url }%
\providecommand \@url [1]{\endgroup\@href {#1}{\urlprefix }}%
\providecommand \urlprefix  [0]{URL }%
\providecommand \Eprint [0]{\href }%
\providecommand \doibase [0]{http://dx.doi.org/}%
\providecommand \selectlanguage [0]{\@gobble}%
\providecommand \bibinfo  [0]{\@secondoftwo}%
\providecommand \bibfield  [0]{\@secondoftwo}%
\providecommand \translation [1]{[#1]}%
\providecommand \BibitemOpen [0]{}%
\providecommand \bibitemStop [0]{}%
\providecommand \bibitemNoStop [0]{.\EOS\space}%
\providecommand \EOS [0]{\spacefactor3000\relax}%
\providecommand \BibitemShut  [1]{\csname bibitem#1\endcsname}%
\let\auto@bib@innerbib\@empty
\bibitem [{\citenamefont {Hasan}\ and\ \citenamefont {Kane}(2010)}]{Hasan_RMP}%
  \BibitemOpen
  \bibfield  {author} {\bibinfo {author} {\bibfnamefont {M.~Z.}\ \bibnamefont
  {Hasan}}\ and\ \bibinfo {author} {\bibfnamefont {C.~L.}\ \bibnamefont
  {Kane}},\ }\href {\doibase 10.1103/RevModPhys.82.3045} {\bibfield  {journal}
  {\bibinfo  {journal} {Rev. Mod. Phys.}\ }\textbf {\bibinfo {volume} {82}},\
  \bibinfo {pages} {3045} (\bibinfo {year} {2010})}\BibitemShut {NoStop}%
\bibitem [{\citenamefont {Qi}\ and\ \citenamefont {Zhang}(2011)}]{Qi_RMP}%
  \BibitemOpen
  \bibfield  {author} {\bibinfo {author} {\bibfnamefont {X.-L.}\ \bibnamefont
  {Qi}}\ and\ \bibinfo {author} {\bibfnamefont {S.-C.}\ \bibnamefont {Zhang}},\
  }\href {\doibase 10.1103/RevModPhys.83.1057} {\bibfield  {journal} {\bibinfo
  {journal} {Rev. Mod. Phys.}\ }\textbf {\bibinfo {volume} {83}},\ \bibinfo
  {pages} {1057} (\bibinfo {year} {2011})}\BibitemShut {NoStop}%
\bibitem [{\citenamefont {Shen}(2012)}]{ShunQingShen_TI}%
  \BibitemOpen
  \bibfield  {author} {\bibinfo {author} {\bibfnamefont {S.-Q.}\ \bibnamefont
  {Shen}},\ }\href {\doibase https://doi.org/10.1007/978-3-642-32858-9} {\emph
  {\bibinfo {title} {Topological Insulators}}}\ (\bibinfo  {publisher}
  {Springer, Berlin},\ \bibinfo {year} {2012})\BibitemShut {NoStop}%
\bibitem [{\citenamefont {Bansil}\ \emph {et~al.}(2016)\citenamefont {Bansil},
  \citenamefont {Lin},\ and\ \citenamefont {Das}}]{Bansil_RMP}%
  \BibitemOpen
  \bibfield  {author} {\bibinfo {author} {\bibfnamefont {A.}~\bibnamefont
  {Bansil}}, \bibinfo {author} {\bibfnamefont {H.}~\bibnamefont {Lin}}, \ and\
  \bibinfo {author} {\bibfnamefont {T.}~\bibnamefont {Das}},\ }\href {\doibase
  10.1103/RevModPhys.88.021004} {\bibfield  {journal} {\bibinfo  {journal}
  {Rev. Mod. Phys.}\ }\textbf {\bibinfo {volume} {88}},\ \bibinfo {pages}
  {021004} (\bibinfo {year} {2016})}\BibitemShut {NoStop}%
\bibitem [{\citenamefont {Zhang}\ \emph {et~al.}(2013)\citenamefont {Zhang},
  \citenamefont {Kane},\ and\ \citenamefont {Mele}}]{ZhangFan_PRL2013}%
  \BibitemOpen
  \bibfield  {author} {\bibinfo {author} {\bibfnamefont {F.}~\bibnamefont
  {Zhang}}, \bibinfo {author} {\bibfnamefont {C.~L.}\ \bibnamefont {Kane}}, \
  and\ \bibinfo {author} {\bibfnamefont {E.~J.}\ \bibnamefont {Mele}},\ }\href
  {\doibase 10.1103/PhysRevLett.110.046404} {\bibfield  {journal} {\bibinfo
  {journal} {Phys. Rev. Lett.}\ }\textbf {\bibinfo {volume} {110}},\ \bibinfo
  {pages} {046404} (\bibinfo {year} {2013})}\BibitemShut {NoStop}%
\bibitem [{\citenamefont {Benalcazar}\ \emph
  {et~al.}(2017{\natexlab{a}})\citenamefont {Benalcazar}, \citenamefont
  {Bernevig},\ and\ \citenamefont {Hughes}}]{Hughes2017}%
  \BibitemOpen
  \bibfield  {author} {\bibinfo {author} {\bibfnamefont {W.~A.}\ \bibnamefont
  {Benalcazar}}, \bibinfo {author} {\bibfnamefont {B.~A.}\ \bibnamefont
  {Bernevig}}, \ and\ \bibinfo {author} {\bibfnamefont {T.~L.}\ \bibnamefont
  {Hughes}},\ }\href {\doibase 10.1126/science.aah6442} {\bibfield  {journal}
  {\bibinfo  {journal} {Science}\ }\textbf {\bibinfo {volume} {357}},\ \bibinfo
  {pages} {61} (\bibinfo {year} {2017}{\natexlab{a}})}\BibitemShut {NoStop}%
\bibitem [{\citenamefont {Langbehn}\ \emph {et~al.}(2017)\citenamefont
  {Langbehn}, \citenamefont {Peng}, \citenamefont {Trifunovic}, \citenamefont
  {von Oppen},\ and\ \citenamefont {Brouwer}}]{Langbehn2017}%
  \BibitemOpen
  \bibfield  {author} {\bibinfo {author} {\bibfnamefont {J.}~\bibnamefont
  {Langbehn}}, \bibinfo {author} {\bibfnamefont {Y.}~\bibnamefont {Peng}},
  \bibinfo {author} {\bibfnamefont {L.}~\bibnamefont {Trifunovic}}, \bibinfo
  {author} {\bibfnamefont {F.}~\bibnamefont {von Oppen}}, \ and\ \bibinfo
  {author} {\bibfnamefont {P.~W.}\ \bibnamefont {Brouwer}},\ }\href {\doibase
  10.1103/PhysRevLett.119.246401} {\bibfield  {journal} {\bibinfo  {journal}
  {Phys. Rev. Lett.}\ }\textbf {\bibinfo {volume} {119}},\ \bibinfo {pages}
  {246401} (\bibinfo {year} {2017})}\BibitemShut {NoStop}%
\bibitem [{\citenamefont {Song}\ \emph {et~al.}(2017)\citenamefont {Song},
  \citenamefont {Fang},\ and\ \citenamefont {Fang}}]{SongZD2017}%
  \BibitemOpen
  \bibfield  {author} {\bibinfo {author} {\bibfnamefont {Z.}~\bibnamefont
  {Song}}, \bibinfo {author} {\bibfnamefont {Z.}~\bibnamefont {Fang}}, \ and\
  \bibinfo {author} {\bibfnamefont {C.}~\bibnamefont {Fang}},\ }\href {\doibase
  10.1103/PhysRevLett.119.246402} {\bibfield  {journal} {\bibinfo  {journal}
  {Phys. Rev. Lett.}\ }\textbf {\bibinfo {volume} {119}},\ \bibinfo {pages}
  {246402} (\bibinfo {year} {2017})}\BibitemShut {NoStop}%
\bibitem [{\citenamefont {Benalcazar}\ \emph
  {et~al.}(2017{\natexlab{b}})\citenamefont {Benalcazar}, \citenamefont
  {Bernevig},\ and\ \citenamefont {Hughes}}]{Hughes2017b}%
  \BibitemOpen
  \bibfield  {author} {\bibinfo {author} {\bibfnamefont {W.~A.}\ \bibnamefont
  {Benalcazar}}, \bibinfo {author} {\bibfnamefont {B.~A.}\ \bibnamefont
  {Bernevig}}, \ and\ \bibinfo {author} {\bibfnamefont {T.~L.}\ \bibnamefont
  {Hughes}},\ }\href {\doibase 10.1103/PhysRevB.96.245115} {\bibfield
  {journal} {\bibinfo  {journal} {Phys. Rev. B}\ }\textbf {\bibinfo {volume}
  {96}},\ \bibinfo {pages} {245115} (\bibinfo {year}
  {2017}{\natexlab{b}})}\BibitemShut {NoStop}%
\bibitem [{\citenamefont {Schindler}\ \emph
  {et~al.}(2018{\natexlab{a}})\citenamefont {Schindler}, \citenamefont {Cook},
  \citenamefont {Vergniory}, \citenamefont {Wang}, \citenamefont {Parkin},
  \citenamefont {Bernevig},\ and\ \citenamefont {Neupert}}]{Schindler2018SA}%
  \BibitemOpen
  \bibfield  {author} {\bibinfo {author} {\bibfnamefont {F.}~\bibnamefont
  {Schindler}}, \bibinfo {author} {\bibfnamefont {A.~M.}\ \bibnamefont {Cook}},
  \bibinfo {author} {\bibfnamefont {M.~G.}\ \bibnamefont {Vergniory}}, \bibinfo
  {author} {\bibfnamefont {Z.}~\bibnamefont {Wang}}, \bibinfo {author}
  {\bibfnamefont {S.~S.~P.}\ \bibnamefont {Parkin}}, \bibinfo {author}
  {\bibfnamefont {B.~A.}\ \bibnamefont {Bernevig}}, \ and\ \bibinfo {author}
  {\bibfnamefont {T.}~\bibnamefont {Neupert}},\ }\href {\doibase
  10.1126/sciadv.aat0346} {\bibfield  {journal} {\bibinfo  {journal} {Science
  Advances}\ }\textbf {\bibinfo {volume} {4}},\ \bibinfo {pages} {eaat0346}
  (\bibinfo {year} {2018}{\natexlab{a}})}\BibitemShut {NoStop}%
\bibitem [{\citenamefont {Bradlyn}\ \emph {et~al.}(2017)\citenamefont
  {Bradlyn}, \citenamefont {Elcoro}, \citenamefont {Cano}, \citenamefont
  {Vergniory}, \citenamefont {Wang}, \citenamefont {Felser}, \citenamefont
  {Aroyo},\ and\ \citenamefont {Bernevig}}]{Bradlyn2017}%
  \BibitemOpen
  \bibfield  {author} {\bibinfo {author} {\bibfnamefont {B.}~\bibnamefont
  {Bradlyn}}, \bibinfo {author} {\bibfnamefont {L.}~\bibnamefont {Elcoro}},
  \bibinfo {author} {\bibfnamefont {J.}~\bibnamefont {Cano}}, \bibinfo {author}
  {\bibfnamefont {M.}~\bibnamefont {Vergniory}}, \bibinfo {author}
  {\bibfnamefont {Z.}~\bibnamefont {Wang}}, \bibinfo {author} {\bibfnamefont
  {C.}~\bibnamefont {Felser}}, \bibinfo {author} {\bibfnamefont {M.~I.}\
  \bibnamefont {Aroyo}}, \ and\ \bibinfo {author} {\bibfnamefont {B.~A.}\
  \bibnamefont {Bernevig}},\ }\href
  {https://www.nature.com/articles/nature23268} {\bibfield  {journal} {\bibinfo
   {journal} {Nature}\ }\textbf {\bibinfo {volume} {547}},\ \bibinfo {pages}
  {298} (\bibinfo {year} {2017})}\BibitemShut {NoStop}%
\bibitem [{\citenamefont {Schindler}\ \emph
  {et~al.}(2018{\natexlab{b}})\citenamefont {Schindler}, \citenamefont {Wang},
  \citenamefont {Vergniory}, \citenamefont {Cook}, \citenamefont {Murani},
  \citenamefont {Sengupta}, \citenamefont {Kasumov}, \citenamefont {Deblock},
  \citenamefont {Jeon}, \citenamefont {Drozdov}, \citenamefont {Bouchiat},
  \citenamefont {Gu{\'e}ron}, \citenamefont {Yazdani}, \citenamefont
  {Bernevig},\ and\ \citenamefont {Neupert}}]{Schindler2018}%
  \BibitemOpen
  \bibfield  {author} {\bibinfo {author} {\bibfnamefont {F.}~\bibnamefont
  {Schindler}}, \bibinfo {author} {\bibfnamefont {Z.}~\bibnamefont {Wang}},
  \bibinfo {author} {\bibfnamefont {M.~G.}\ \bibnamefont {Vergniory}}, \bibinfo
  {author} {\bibfnamefont {A.~M.}\ \bibnamefont {Cook}}, \bibinfo {author}
  {\bibfnamefont {A.}~\bibnamefont {Murani}}, \bibinfo {author} {\bibfnamefont
  {S.}~\bibnamefont {Sengupta}}, \bibinfo {author} {\bibfnamefont {A.~Y.}\
  \bibnamefont {Kasumov}}, \bibinfo {author} {\bibfnamefont {R.}~\bibnamefont
  {Deblock}}, \bibinfo {author} {\bibfnamefont {S.}~\bibnamefont {Jeon}},
  \bibinfo {author} {\bibfnamefont {I.}~\bibnamefont {Drozdov}}, \bibinfo
  {author} {\bibfnamefont {H.}~\bibnamefont {Bouchiat}}, \bibinfo {author}
  {\bibfnamefont {S.}~\bibnamefont {Gu{\'e}ron}}, \bibinfo {author}
  {\bibfnamefont {A.}~\bibnamefont {Yazdani}}, \bibinfo {author} {\bibfnamefont
  {B.~A.}\ \bibnamefont {Bernevig}}, \ and\ \bibinfo {author} {\bibfnamefont
  {T.}~\bibnamefont {Neupert}},\ }\href {\doibase
  https://doi.org/10.1038/s41567-018-0224-7} {\bibfield  {journal} {\bibinfo
  {journal} {Nat. Phys.}\ }\textbf {\bibinfo {volume} {14}},\ \bibinfo {pages}
  {918} (\bibinfo {year} {2018}{\natexlab{b}})}\BibitemShut {NoStop}%
\bibitem [{\citenamefont {Yue}\ \emph {et~al.}(2019)\citenamefont {Yue},
  \citenamefont {Xu}, \citenamefont {Song}, \citenamefont {Weng}, \citenamefont
  {Lu}, \citenamefont {Fang},\ and\ \citenamefont {Dai}}]{Yue2019ws}%
  \BibitemOpen
  \bibfield  {author} {\bibinfo {author} {\bibfnamefont {C.}~\bibnamefont
  {Yue}}, \bibinfo {author} {\bibfnamefont {Y.}~\bibnamefont {Xu}}, \bibinfo
  {author} {\bibfnamefont {Z.}~\bibnamefont {Song}}, \bibinfo {author}
  {\bibfnamefont {H.}~\bibnamefont {Weng}}, \bibinfo {author} {\bibfnamefont
  {Y.-M.}\ \bibnamefont {Lu}}, \bibinfo {author} {\bibfnamefont
  {C.}~\bibnamefont {Fang}}, \ and\ \bibinfo {author} {\bibfnamefont
  {X.}~\bibnamefont {Dai}},\ }\href {\doibase
  https://doi.org/10.1038/s41567-019-0457-0} {\bibfield  {journal} {\bibinfo
  {journal} {Nature Physics}\ }\textbf {\bibinfo {volume} {15}},\ \bibinfo
  {pages} {577} (\bibinfo {year} {2019})}\BibitemShut {NoStop}%
\bibitem [{\citenamefont {Wang}\ \emph {et~al.}(2019)\citenamefont {Wang},
  \citenamefont {Wieder}, \citenamefont {Li}, \citenamefont {Yan},\ and\
  \citenamefont {Bernevig}}]{WangZhijun_arXiv}%
  \BibitemOpen
  \bibfield  {author} {\bibinfo {author} {\bibfnamefont {Z.}~\bibnamefont
  {Wang}}, \bibinfo {author} {\bibfnamefont {B.~J.}\ \bibnamefont {Wieder}},
  \bibinfo {author} {\bibfnamefont {J.}~\bibnamefont {Li}}, \bibinfo {author}
  {\bibfnamefont {B.}~\bibnamefont {Yan}}, \ and\ \bibinfo {author}
  {\bibfnamefont {B.~A.}\ \bibnamefont {Bernevig}},\ }\href {\doibase
  10.1103/PhysRevLett.123.186401} {\bibfield  {journal} {\bibinfo  {journal}
  {Phys. Rev. Lett.}\ }\textbf {\bibinfo {volume} {123}},\ \bibinfo {pages}
  {186401} (\bibinfo {year} {2019})}\BibitemShut {NoStop}%
\bibitem [{\citenamefont {Zhang}\ \emph {et~al.}(2019)\citenamefont {Zhang},
  \citenamefont {Jiang}, \citenamefont {Song}, \citenamefont {Huang},
  \citenamefont {He}, \citenamefont {Fang}, \citenamefont {Weng},\ and\
  \citenamefont {Fang}}]{Zhang2019tp}%
  \BibitemOpen
  \bibfield  {author} {\bibinfo {author} {\bibfnamefont {T.}~\bibnamefont
  {Zhang}}, \bibinfo {author} {\bibfnamefont {Y.}~\bibnamefont {Jiang}},
  \bibinfo {author} {\bibfnamefont {Z.}~\bibnamefont {Song}}, \bibinfo {author}
  {\bibfnamefont {H.}~\bibnamefont {Huang}}, \bibinfo {author} {\bibfnamefont
  {Y.}~\bibnamefont {He}}, \bibinfo {author} {\bibfnamefont {Z.}~\bibnamefont
  {Fang}}, \bibinfo {author} {\bibfnamefont {H.}~\bibnamefont {Weng}}, \ and\
  \bibinfo {author} {\bibfnamefont {C.}~\bibnamefont {Fang}},\ }\href
  {https://www.nature.com/articles/s41586-019-0944-6} {\bibfield  {journal}
  {\bibinfo  {journal} {Nature}\ }\textbf {\bibinfo {volume} {566}},\ \bibinfo
  {pages} {475} (\bibinfo {year} {2019})}\BibitemShut {NoStop}%
\bibitem [{\citenamefont {Vergniory}\ \emph {et~al.}(2019)\citenamefont
  {Vergniory}, \citenamefont {Elcoro}, \citenamefont {Felser}, \citenamefont
  {Regnault}, \citenamefont {Bernevig},\ and\ \citenamefont
  {Wang}}]{Vergniory2019ub}%
  \BibitemOpen
  \bibfield  {author} {\bibinfo {author} {\bibfnamefont {M.}~\bibnamefont
  {Vergniory}}, \bibinfo {author} {\bibfnamefont {L.}~\bibnamefont {Elcoro}},
  \bibinfo {author} {\bibfnamefont {C.}~\bibnamefont {Felser}}, \bibinfo
  {author} {\bibfnamefont {N.}~\bibnamefont {Regnault}}, \bibinfo {author}
  {\bibfnamefont {B.~A.}\ \bibnamefont {Bernevig}}, \ and\ \bibinfo {author}
  {\bibfnamefont {Z.}~\bibnamefont {Wang}},\ }\href
  {https://www.nature.com/articles/s41586-019-0954-4} {\bibfield  {journal}
  {\bibinfo  {journal} {Nature}\ }\textbf {\bibinfo {volume} {566}},\ \bibinfo
  {pages} {480} (\bibinfo {year} {2019})}\BibitemShut {NoStop}%
\bibitem [{\citenamefont {Tang}\ \emph
  {et~al.}(2019{\natexlab{a}})\citenamefont {Tang}, \citenamefont {Po},
  \citenamefont {Vishwanath},\ and\ \citenamefont {Wan}}]{TangNP}%
  \BibitemOpen
  \bibfield  {author} {\bibinfo {author} {\bibfnamefont {F.}~\bibnamefont
  {Tang}}, \bibinfo {author} {\bibfnamefont {H.~C.}\ \bibnamefont {Po}},
  \bibinfo {author} {\bibfnamefont {A.}~\bibnamefont {Vishwanath}}, \ and\
  \bibinfo {author} {\bibfnamefont {X.}~\bibnamefont {Wan}},\ }\href
  {https://www.nature.com/articles/s41567-019-0418-7} {\bibfield  {journal}
  {\bibinfo  {journal} {Nature Physics}\ }\textbf {\bibinfo {volume} {15}},\
  \bibinfo {pages} {470} (\bibinfo {year} {2019}{\natexlab{a}})}\BibitemShut
  {NoStop}%
\bibitem [{\citenamefont {Tang}\ \emph
  {et~al.}(2019{\natexlab{b}})\citenamefont {Tang}, \citenamefont {Po},
  \citenamefont {Vishwanath},\ and\ \citenamefont {Wan}}]{TangNature}%
  \BibitemOpen
  \bibfield  {author} {\bibinfo {author} {\bibfnamefont {F.}~\bibnamefont
  {Tang}}, \bibinfo {author} {\bibfnamefont {H.~C.}\ \bibnamefont {Po}},
  \bibinfo {author} {\bibfnamefont {A.}~\bibnamefont {Vishwanath}}, \ and\
  \bibinfo {author} {\bibfnamefont {X.}~\bibnamefont {Wan}},\ }\href
  {https://www.nature.com/articles/s41586-019-0937-5} {\bibfield  {journal}
  {\bibinfo  {journal} {Nature}\ }\textbf {\bibinfo {volume} {566}},\ \bibinfo
  {pages} {486} (\bibinfo {year} {2019}{\natexlab{b}})}\BibitemShut {NoStop}%
\bibitem [{\citenamefont {Xu}\ \emph {et~al.}(2019)\citenamefont {Xu},
  \citenamefont {Song}, \citenamefont {Wang}, \citenamefont {Weng},\ and\
  \citenamefont {Dai}}]{XuYF2019}%
  \BibitemOpen
  \bibfield  {author} {\bibinfo {author} {\bibfnamefont {Y.}~\bibnamefont
  {Xu}}, \bibinfo {author} {\bibfnamefont {Z.}~\bibnamefont {Song}}, \bibinfo
  {author} {\bibfnamefont {Z.}~\bibnamefont {Wang}}, \bibinfo {author}
  {\bibfnamefont {H.}~\bibnamefont {Weng}}, \ and\ \bibinfo {author}
  {\bibfnamefont {X.}~\bibnamefont {Dai}},\ }\href {\doibase
  10.1103/PhysRevLett.122.256402} {\bibfield  {journal} {\bibinfo  {journal}
  {Phys. Rev. Lett.}\ }\textbf {\bibinfo {volume} {122}},\ \bibinfo {pages}
  {256402} (\bibinfo {year} {2019})}\BibitemShut {NoStop}%
\bibitem [{\citenamefont {Sheng}\ \emph {et~al.}(2019)\citenamefont {Sheng},
  \citenamefont {Chen}, \citenamefont {Liu}, \citenamefont {Chen},
  \citenamefont {Yu}, \citenamefont {Zhao},\ and\ \citenamefont
  {Yang}}]{Sheng2019}%
  \BibitemOpen
  \bibfield  {author} {\bibinfo {author} {\bibfnamefont {X.-L.}\ \bibnamefont
  {Sheng}}, \bibinfo {author} {\bibfnamefont {C.}~\bibnamefont {Chen}},
  \bibinfo {author} {\bibfnamefont {H.}~\bibnamefont {Liu}}, \bibinfo {author}
  {\bibfnamefont {Z.}~\bibnamefont {Chen}}, \bibinfo {author} {\bibfnamefont
  {Z.-M.}\ \bibnamefont {Yu}}, \bibinfo {author} {\bibfnamefont {Y.~X.}\
  \bibnamefont {Zhao}}, \ and\ \bibinfo {author} {\bibfnamefont {S.~A.}\
  \bibnamefont {Yang}},\ }\href {\doibase 10.1103/PhysRevLett.123.256402}
  {\bibfield  {journal} {\bibinfo  {journal} {Phys. Rev. Lett.}\ }\textbf
  {\bibinfo {volume} {123}},\ \bibinfo {pages} {256402} (\bibinfo {year}
  {2019})}\BibitemShut {NoStop}%
\bibitem [{\citenamefont {Liu}\ \emph {et~al.}(2019)\citenamefont {Liu},
  \citenamefont {Zhao}, \citenamefont {Liu},\ and\ \citenamefont
  {Wang}}]{WangZF2019}%
  \BibitemOpen
  \bibfield  {author} {\bibinfo {author} {\bibfnamefont {B.}~\bibnamefont
  {Liu}}, \bibinfo {author} {\bibfnamefont {G.}~\bibnamefont {Zhao}}, \bibinfo
  {author} {\bibfnamefont {Z.}~\bibnamefont {Liu}}, \ and\ \bibinfo {author}
  {\bibfnamefont {Z.~F.}\ \bibnamefont {Wang}},\ }\href {\doibase
  10.1021/acs.nanolett.9b02719} {\bibfield  {journal} {\bibinfo  {journal}
  {Nano Letters}\ }\textbf {\bibinfo {volume} {19}},\ \bibinfo {pages} {6492}
  (\bibinfo {year} {2019})}\BibitemShut {NoStop}%
\bibitem [{\citenamefont {Lee}\ \emph {et~al.}(2020)\citenamefont {Lee},
  \citenamefont {Kim}, \citenamefont {Ahn},\ and\ \citenamefont
  {Yang}}]{Lee2020th}%
  \BibitemOpen
  \bibfield  {author} {\bibinfo {author} {\bibfnamefont {E.}~\bibnamefont
  {Lee}}, \bibinfo {author} {\bibfnamefont {R.}~\bibnamefont {Kim}}, \bibinfo
  {author} {\bibfnamefont {J.}~\bibnamefont {Ahn}}, \ and\ \bibinfo {author}
  {\bibfnamefont {B.-J.}\ \bibnamefont {Yang}},\ }\href@noop {} {\bibfield
  {journal} {\bibinfo  {journal} {npj Quantum Materials}\ }\textbf {\bibinfo
  {volume} {5}},\ \bibinfo {pages} {1} (\bibinfo {year} {2020})}\BibitemShut
  {NoStop}%
\bibitem [{\citenamefont {Chen}\ \emph {et~al.}(2021)\citenamefont {Chen},
  \citenamefont {Wu}, \citenamefont {Yu}, \citenamefont {Chen}, \citenamefont
  {Zhao}, \citenamefont {Sheng},\ and\ \citenamefont {Yang}}]{chen2020GPY}%
  \BibitemOpen
  \bibfield  {author} {\bibinfo {author} {\bibfnamefont {C.}~\bibnamefont
  {Chen}}, \bibinfo {author} {\bibfnamefont {W.}~\bibnamefont {Wu}}, \bibinfo
  {author} {\bibfnamefont {Z.-M.}\ \bibnamefont {Yu}}, \bibinfo {author}
  {\bibfnamefont {Z.}~\bibnamefont {Chen}}, \bibinfo {author} {\bibfnamefont
  {Y.~X.}\ \bibnamefont {Zhao}}, \bibinfo {author} {\bibfnamefont {X.-L.}\
  \bibnamefont {Sheng}}, \ and\ \bibinfo {author} {\bibfnamefont {S.~A.}\
  \bibnamefont {Yang}},\ }\href {\doibase 10.1103/PhysRevB.104.085205}
  {\bibfield  {journal} {\bibinfo  {journal} {Phys. Rev. B}\ }\textbf {\bibinfo
  {volume} {104}},\ \bibinfo {pages} {085205} (\bibinfo {year}
  {2021})}\BibitemShut {NoStop}%
\bibitem [{\citenamefont {Park}\ \emph {et~al.}(2019)\citenamefont {Park},
  \citenamefont {Kim}, \citenamefont {Cho},\ and\ \citenamefont
  {Lee}}]{Park2019}%
  \BibitemOpen
  \bibfield  {author} {\bibinfo {author} {\bibfnamefont {M.~J.}\ \bibnamefont
  {Park}}, \bibinfo {author} {\bibfnamefont {Y.}~\bibnamefont {Kim}}, \bibinfo
  {author} {\bibfnamefont {G.~Y.}\ \bibnamefont {Cho}}, \ and\ \bibinfo
  {author} {\bibfnamefont {S.~B.}\ \bibnamefont {Lee}},\ }\href {\doibase
  10.1103/PhysRevLett.123.216803} {\bibfield  {journal} {\bibinfo  {journal}
  {Phys. Rev. Lett.}\ }\textbf {\bibinfo {volume} {123}},\ \bibinfo {pages}
  {216803} (\bibinfo {year} {2019})}\BibitemShut {NoStop}%
\bibitem [{\citenamefont {Liu}\ \emph {et~al.}(2021)\citenamefont {Liu},
  \citenamefont {Xian}, \citenamefont {Mu}, \citenamefont {Zhao}, \citenamefont
  {Liu}, \citenamefont {Rubio},\ and\ \citenamefont {Wang}}]{Liu2021}%
  \BibitemOpen
  \bibfield  {author} {\bibinfo {author} {\bibfnamefont {B.}~\bibnamefont
  {Liu}}, \bibinfo {author} {\bibfnamefont {L.}~\bibnamefont {Xian}}, \bibinfo
  {author} {\bibfnamefont {H.}~\bibnamefont {Mu}}, \bibinfo {author}
  {\bibfnamefont {G.}~\bibnamefont {Zhao}}, \bibinfo {author} {\bibfnamefont
  {Z.}~\bibnamefont {Liu}}, \bibinfo {author} {\bibfnamefont {A.}~\bibnamefont
  {Rubio}}, \ and\ \bibinfo {author} {\bibfnamefont {Z.~F.}\ \bibnamefont
  {Wang}},\ }\href {\doibase 10.1103/PhysRevLett.126.066401} {\bibfield
  {journal} {\bibinfo  {journal} {Phys. Rev. Lett.}\ }\textbf {\bibinfo
  {volume} {126}},\ \bibinfo {pages} {066401} (\bibinfo {year}
  {2021})}\BibitemShut {NoStop}%
\bibitem [{\citenamefont {Ezawa}(2018)}]{Ezawa2018P}%
  \BibitemOpen
  \bibfield  {author} {\bibinfo {author} {\bibfnamefont {M.}~\bibnamefont
  {Ezawa}},\ }\href {\doibase 10.1103/PhysRevB.98.045125} {\bibfield  {journal}
  {\bibinfo  {journal} {Phys. Rev. B}\ }\textbf {\bibinfo {volume} {98}},\
  \bibinfo {pages} {045125} (\bibinfo {year} {2018})}\BibitemShut {NoStop}%
\bibitem [{\citenamefont {Hitomi}\ \emph {et~al.}(2021)\citenamefont {Hitomi},
  \citenamefont {Kawakami},\ and\ \citenamefont {Koshino}}]{Hitomi2021}%
  \BibitemOpen
  \bibfield  {author} {\bibinfo {author} {\bibfnamefont {M.}~\bibnamefont
  {Hitomi}}, \bibinfo {author} {\bibfnamefont {T.}~\bibnamefont {Kawakami}}, \
  and\ \bibinfo {author} {\bibfnamefont {M.}~\bibnamefont {Koshino}},\ }\href
  {\doibase 10.1103/PhysRevB.104.125302} {\bibfield  {journal} {\bibinfo
  {journal} {Phys. Rev. B}\ }\textbf {\bibinfo {volume} {104}},\ \bibinfo
  {pages} {125302} (\bibinfo {year} {2021})}\BibitemShut {NoStop}%
\bibitem [{\citenamefont {Chen}\ \emph {et~al.}(2020)\citenamefont {Chen},
  \citenamefont {Song}, \citenamefont {Zhao}, \citenamefont {Chen},
  \citenamefont {Yu}, \citenamefont {Sheng},\ and\ \citenamefont
  {Yang}}]{Chen2020}%
  \BibitemOpen
  \bibfield  {author} {\bibinfo {author} {\bibfnamefont {C.}~\bibnamefont
  {Chen}}, \bibinfo {author} {\bibfnamefont {Z.}~\bibnamefont {Song}}, \bibinfo
  {author} {\bibfnamefont {J.-Z.}\ \bibnamefont {Zhao}}, \bibinfo {author}
  {\bibfnamefont {Z.}~\bibnamefont {Chen}}, \bibinfo {author} {\bibfnamefont
  {Z.-M.}\ \bibnamefont {Yu}}, \bibinfo {author} {\bibfnamefont {X.-L.}\
  \bibnamefont {Sheng}}, \ and\ \bibinfo {author} {\bibfnamefont {S.~A.}\
  \bibnamefont {Yang}},\ }\href {\doibase 10.1103/PhysRevLett.125.056402}
  {\bibfield  {journal} {\bibinfo  {journal} {Phys. Rev. Lett.}\ }\textbf
  {\bibinfo {volume} {125}},\ \bibinfo {pages} {056402} (\bibinfo {year}
  {2020})}\BibitemShut {NoStop}%
\bibitem [{\citenamefont {Radha}\ and\ \citenamefont
  {Lambrecht}(2020)}]{Radha2020}%
  \BibitemOpen
  \bibfield  {author} {\bibinfo {author} {\bibfnamefont {S.~K.}\ \bibnamefont
  {Radha}}\ and\ \bibinfo {author} {\bibfnamefont {W.~R.~L.}\ \bibnamefont
  {Lambrecht}},\ }\href {\doibase 10.1103/PhysRevB.102.115104} {\bibfield
  {journal} {\bibinfo  {journal} {Phys. Rev. B}\ }\textbf {\bibinfo {volume}
  {102}},\ \bibinfo {pages} {115104} (\bibinfo {year} {2020})}\BibitemShut
  {NoStop}%
\bibitem [{\citenamefont {Huang}\ and\ \citenamefont {Liu}(2021)}]{Huang2021}%
  \BibitemOpen
  \bibfield  {author} {\bibinfo {author} {\bibfnamefont {H.}~\bibnamefont
  {Huang}}\ and\ \bibinfo {author} {\bibfnamefont {F.}~\bibnamefont {Liu}},\
  }\href {\doibase https://doi.org/10.1093/nsr/nwab170} {\bibfield  {journal}
  {\bibinfo  {journal} {National Science Review}\ } (\bibinfo {year} {2021}),\
  https://doi.org/10.1093/nsr/nwab170}\BibitemShut {NoStop}%
\bibitem [{\citenamefont {Qian}\ \emph {et~al.}(2021)\citenamefont {Qian},
  \citenamefont {Liu},\ and\ \citenamefont {Yao}}]{Qian2021}%
  \BibitemOpen
  \bibfield  {author} {\bibinfo {author} {\bibfnamefont {S.}~\bibnamefont
  {Qian}}, \bibinfo {author} {\bibfnamefont {C.-C.}\ \bibnamefont {Liu}}, \
  and\ \bibinfo {author} {\bibfnamefont {Y.}~\bibnamefont {Yao}},\ }\href
  {\doibase 10.1103/PhysRevB.104.245427} {\bibfield  {journal} {\bibinfo
  {journal} {Phys. Rev. B}\ }\textbf {\bibinfo {volume} {104}},\ \bibinfo
  {pages} {245427} (\bibinfo {year} {2021})}\BibitemShut {NoStop}%
\bibitem [{\citenamefont {Pan}\ \emph {et~al.}(2022)\citenamefont {Pan},
  \citenamefont {Li}, \citenamefont {Fan},\ and\ \citenamefont
  {Huang}}]{Pan2022}%
  \BibitemOpen
  \bibfield  {author} {\bibinfo {author} {\bibfnamefont {M.}~\bibnamefont
  {Pan}}, \bibinfo {author} {\bibfnamefont {D.}~\bibnamefont {Li}}, \bibinfo
  {author} {\bibfnamefont {J.}~\bibnamefont {Fan}}, \ and\ \bibinfo {author}
  {\bibfnamefont {H.}~\bibnamefont {Huang}},\ }\href {\doibase
  10.1038/s41524-021-00695-2} {\bibfield  {journal} {\bibinfo  {journal} {npj
  Computational Materials}\ }\textbf {\bibinfo {volume} {8}},\ \bibinfo {pages}
  {1} (\bibinfo {year} {2022})}\BibitemShut {NoStop}%
\bibitem [{\citenamefont {Zhao}\ and\ \citenamefont {Lu}(2017)}]{RCN2017}%
  \BibitemOpen
  \bibfield  {author} {\bibinfo {author} {\bibfnamefont {Y.~X.}\ \bibnamefont
  {Zhao}}\ and\ \bibinfo {author} {\bibfnamefont {Y.}~\bibnamefont {Lu}},\
  }\href {\doibase 10.1103/PhysRevLett.118.056401} {\bibfield  {journal}
  {\bibinfo  {journal} {Phys. Rev. Lett.}\ }\textbf {\bibinfo {volume} {118}},\
  \bibinfo {pages} {056401} (\bibinfo {year} {2017})}\BibitemShut {NoStop}%
\bibitem [{\citenamefont {Nakahara}(2003)}]{Nakahara2003}%
  \BibitemOpen
  \bibfield  {author} {\bibinfo {author} {\bibfnamefont {M.}~\bibnamefont
  {Nakahara}},\ }\href@noop {} {\emph {\bibinfo {title} {Geometry, Topology and
  Physics, 2nd ed.}}}\ (\bibinfo  {publisher} {Institute of Physics, Bristol},\
  \bibinfo {year} {2003})\BibitemShut {NoStop}%
\bibitem [{\citenamefont {Song}\ \emph {et~al.}(2019)\citenamefont {Song},
  \citenamefont {Wang}, \citenamefont {Shi}, \citenamefont {Li}, \citenamefont
  {Fang},\ and\ \citenamefont {Bernevig}}]{Song2019}%
  \BibitemOpen
  \bibfield  {author} {\bibinfo {author} {\bibfnamefont {Z.}~\bibnamefont
  {Song}}, \bibinfo {author} {\bibfnamefont {Z.}~\bibnamefont {Wang}}, \bibinfo
  {author} {\bibfnamefont {W.}~\bibnamefont {Shi}}, \bibinfo {author}
  {\bibfnamefont {G.}~\bibnamefont {Li}}, \bibinfo {author} {\bibfnamefont
  {C.}~\bibnamefont {Fang}}, \ and\ \bibinfo {author} {\bibfnamefont {B.~A.}\
  \bibnamefont {Bernevig}},\ }\href {\doibase 10.1103/PhysRevLett.123.036401}
  {\bibfield  {journal} {\bibinfo  {journal} {Phys. Rev. Lett.}\ }\textbf
  {\bibinfo {volume} {123}},\ \bibinfo {pages} {036401} (\bibinfo {year}
  {2019})}\BibitemShut {NoStop}%
\bibitem [{\citenamefont {Ahn}\ \emph {et~al.}(2019)\citenamefont {Ahn},
  \citenamefont {Park},\ and\ \citenamefont {Yang}}]{Ahn2019}%
  \BibitemOpen
  \bibfield  {author} {\bibinfo {author} {\bibfnamefont {J.}~\bibnamefont
  {Ahn}}, \bibinfo {author} {\bibfnamefont {S.}~\bibnamefont {Park}}, \ and\
  \bibinfo {author} {\bibfnamefont {B.-J.}\ \bibnamefont {Yang}},\ }\href
  {\doibase 10.1103/PhysRevX.9.021013} {\bibfield  {journal} {\bibinfo
  {journal} {Phys. Rev. X}\ }\textbf {\bibinfo {volume} {9}},\ \bibinfo {pages}
  {021013} (\bibinfo {year} {2019})}\BibitemShut {NoStop}%
\bibitem [{\citenamefont {Wang}\ \emph {et~al.}(2020)\citenamefont {Wang},
  \citenamefont {Dai}, \citenamefont {Shao}, \citenamefont {Yang},\ and\
  \citenamefont {Zhao}}]{Zhao2020}%
  \BibitemOpen
  \bibfield  {author} {\bibinfo {author} {\bibfnamefont {K.}~\bibnamefont
  {Wang}}, \bibinfo {author} {\bibfnamefont {J.-X.}\ \bibnamefont {Dai}},
  \bibinfo {author} {\bibfnamefont {L.~B.}\ \bibnamefont {Shao}}, \bibinfo
  {author} {\bibfnamefont {S.~A.}\ \bibnamefont {Yang}}, \ and\ \bibinfo
  {author} {\bibfnamefont {Y.~X.}\ \bibnamefont {Zhao}},\ }\href {\doibase
  10.1103/PhysRevLett.125.126403} {\bibfield  {journal} {\bibinfo  {journal}
  {Phys. Rev. Lett.}\ }\textbf {\bibinfo {volume} {125}},\ \bibinfo {pages}
  {126403} (\bibinfo {year} {2020})}\BibitemShut {NoStop}%
\bibitem [{\citenamefont {Baughman}\ \emph {et~al.}(1987)\citenamefont
  {Baughman}, \citenamefont {Eckhardt},\ and\ \citenamefont
  {Kertesz}}]{Baughman1987}%
  \BibitemOpen
  \bibfield  {author} {\bibinfo {author} {\bibfnamefont {R.~H.}\ \bibnamefont
  {Baughman}}, \bibinfo {author} {\bibfnamefont {H.}~\bibnamefont {Eckhardt}},
  \ and\ \bibinfo {author} {\bibfnamefont {M.}~\bibnamefont {Kertesz}},\ }\href
  {\doibase https://doi.org/10.1063/1.453405} {\bibfield  {journal} {\bibinfo
  {journal} {The Journal of Chemical Physics}\ }\textbf {\bibinfo {volume}
  {87}},\ \bibinfo {pages} {6687} (\bibinfo {year} {1987})}\BibitemShut
  {NoStop}%
\bibitem [{\citenamefont {Malko}\ \emph {et~al.}(2012)\citenamefont {Malko},
  \citenamefont {Neiss}, \citenamefont {Vi\~nes},\ and\ \citenamefont
  {G\"orling}}]{GPY2012}%
  \BibitemOpen
  \bibfield  {author} {\bibinfo {author} {\bibfnamefont {D.}~\bibnamefont
  {Malko}}, \bibinfo {author} {\bibfnamefont {C.}~\bibnamefont {Neiss}},
  \bibinfo {author} {\bibfnamefont {F.}~\bibnamefont {Vi\~nes}}, \ and\
  \bibinfo {author} {\bibfnamefont {A.}~\bibnamefont {G\"orling}},\ }\href
  {\doibase 10.1103/PhysRevLett.108.086804} {\bibfield  {journal} {\bibinfo
  {journal} {Phys. Rev. Lett.}\ }\textbf {\bibinfo {volume} {108}},\ \bibinfo
  {pages} {086804} (\bibinfo {year} {2012})}\BibitemShut {NoStop}%
\bibitem [{\citenamefont {Li}\ \emph {et~al.}(2010)\citenamefont {Li},
  \citenamefont {Li}, \citenamefont {Liu}, \citenamefont {Guo}, \citenamefont
  {Li},\ and\ \citenamefont {Zhu}}]{LiYL2010}%
  \BibitemOpen
  \bibfield  {author} {\bibinfo {author} {\bibfnamefont {G.}~\bibnamefont
  {Li}}, \bibinfo {author} {\bibfnamefont {Y.}~\bibnamefont {Li}}, \bibinfo
  {author} {\bibfnamefont {H.}~\bibnamefont {Liu}}, \bibinfo {author}
  {\bibfnamefont {Y.}~\bibnamefont {Guo}}, \bibinfo {author} {\bibfnamefont
  {Y.}~\bibnamefont {Li}}, \ and\ \bibinfo {author} {\bibfnamefont
  {D.}~\bibnamefont {Zhu}},\ }\href {\doibase 10.1039/B922733D} {\bibfield
  {journal} {\bibinfo  {journal} {Chem. Commun.}\ }\textbf {\bibinfo {volume}
  {46}},\ \bibinfo {pages} {3256} (\bibinfo {year} {2010})}\BibitemShut
  {NoStop}%
\bibitem [{\citenamefont {Chen}\ \emph {et~al.}(2017)\citenamefont {Chen},
  \citenamefont {Molina-Jir{\'o}n}, \citenamefont {Klyatskaya}, \citenamefont
  {Klappenberger},\ and\ \citenamefont {Ruben}}]{GDYreview2017}%
  \BibitemOpen
  \bibfield  {author} {\bibinfo {author} {\bibfnamefont {Z.}~\bibnamefont
  {Chen}}, \bibinfo {author} {\bibfnamefont {C.}~\bibnamefont
  {Molina-Jir{\'o}n}}, \bibinfo {author} {\bibfnamefont {S.}~\bibnamefont
  {Klyatskaya}}, \bibinfo {author} {\bibfnamefont {F.}~\bibnamefont
  {Klappenberger}}, \ and\ \bibinfo {author} {\bibfnamefont {M.}~\bibnamefont
  {Ruben}},\ }\href {\doibase https://doi.org/10.1002/andp.201700056}
  {\bibfield  {journal} {\bibinfo  {journal} {Annalen der Physik}\ }\textbf
  {\bibinfo {volume} {529}},\ \bibinfo {pages} {1700056} (\bibinfo {year}
  {2017})}\BibitemShut {NoStop}%
\bibitem [{\citenamefont {Jia}\ \emph {et~al.}(2017)\citenamefont {Jia},
  \citenamefont {Li}, \citenamefont {Zuo}, \citenamefont {Liu}, \citenamefont
  {Huang},\ and\ \citenamefont {Li}}]{LiYL2017}%
  \BibitemOpen
  \bibfield  {author} {\bibinfo {author} {\bibfnamefont {Z.}~\bibnamefont
  {Jia}}, \bibinfo {author} {\bibfnamefont {Y.}~\bibnamefont {Li}}, \bibinfo
  {author} {\bibfnamefont {Z.}~\bibnamefont {Zuo}}, \bibinfo {author}
  {\bibfnamefont {H.}~\bibnamefont {Liu}}, \bibinfo {author} {\bibfnamefont
  {C.}~\bibnamefont {Huang}}, \ and\ \bibinfo {author} {\bibfnamefont
  {Y.}~\bibnamefont {Li}},\ }\href@noop {} {\bibfield  {journal} {\bibinfo
  {journal} {Accounts of Chemical Research}\ }\textbf {\bibinfo {volume}
  {50}},\ \bibinfo {pages} {2470} (\bibinfo {year} {2017})}\BibitemShut
  {NoStop}%
\bibitem [{\citenamefont {Gao}\ \emph {et~al.}(2019)\citenamefont {Gao},
  \citenamefont {Liu}, \citenamefont {Wang},\ and\ \citenamefont
  {Zhang}}]{ZhangJ2019}%
  \BibitemOpen
  \bibfield  {author} {\bibinfo {author} {\bibfnamefont {X.}~\bibnamefont
  {Gao}}, \bibinfo {author} {\bibfnamefont {H.}~\bibnamefont {Liu}}, \bibinfo
  {author} {\bibfnamefont {D.}~\bibnamefont {Wang}}, \ and\ \bibinfo {author}
  {\bibfnamefont {J.}~\bibnamefont {Zhang}},\ }\href {\doibase
  10.1039/C8CS00773J} {\bibfield  {journal} {\bibinfo  {journal} {Chem. Soc.
  Rev.}\ }\textbf {\bibinfo {volume} {48}},\ \bibinfo {pages} {908} (\bibinfo
  {year} {2019})}\BibitemShut {NoStop}%
\bibitem [{\citenamefont {Jana}\ \emph {et~al.}(2019)\citenamefont {Jana},
  \citenamefont {Bandyopadhyay},\ and\ \citenamefont {Jana}}]{Jana2019}%
  \BibitemOpen
  \bibfield  {author} {\bibinfo {author} {\bibfnamefont {S.}~\bibnamefont
  {Jana}}, \bibinfo {author} {\bibfnamefont {A.}~\bibnamefont {Bandyopadhyay}},
  \ and\ \bibinfo {author} {\bibfnamefont {D.}~\bibnamefont {Jana}},\ }\href
  {\doibase 10.1039/C9CP01914F} {\bibfield  {journal} {\bibinfo  {journal}
  {Phys. Chem. Chem. Phys.}\ }\textbf {\bibinfo {volume} {21}},\ \bibinfo
  {pages} {13795} (\bibinfo {year} {2019})}\BibitemShut {NoStop}%
\bibitem [{\citenamefont {Kim}\ and\ \citenamefont {Choi}(2012)}]{Kim2012}%
  \BibitemOpen
  \bibfield  {author} {\bibinfo {author} {\bibfnamefont {B.~G.}\ \bibnamefont
  {Kim}}\ and\ \bibinfo {author} {\bibfnamefont {H.~J.}\ \bibnamefont {Choi}},\
  }\href {\doibase 10.1103/PhysRevB.86.115435} {\bibfield  {journal} {\bibinfo
  {journal} {Phys. Rev. B}\ }\textbf {\bibinfo {volume} {86}},\ \bibinfo
  {pages} {115435} (\bibinfo {year} {2012})}\BibitemShut {NoStop}%
\bibitem [{\citenamefont {Bandyopadhyay}\ and\ \citenamefont
  {Jana}(2020)}]{Band2020}%
  \BibitemOpen
  \bibfield  {author} {\bibinfo {author} {\bibfnamefont {A.}~\bibnamefont
  {Bandyopadhyay}}\ and\ \bibinfo {author} {\bibfnamefont {D.}~\bibnamefont
  {Jana}},\ }\href {\doibase 10.1088/1361-6633/ab85ba} {\bibfield  {journal}
  {\bibinfo  {journal} {Reports on Progress in Physics}\ }\textbf {\bibinfo
  {volume} {83}},\ \bibinfo {pages} {056501} (\bibinfo {year}
  {2020})}\BibitemShut {NoStop}%
\bibitem [{\citenamefont {Jana}\ \emph {et~al.}(2021)\citenamefont {Jana},
  \citenamefont {Bandyopadhyay}, \citenamefont {Datta}, \citenamefont
  {Bhattacharya},\ and\ \citenamefont {Jana}}]{Jana2021}%
  \BibitemOpen
  \bibfield  {author} {\bibinfo {author} {\bibfnamefont {S.}~\bibnamefont
  {Jana}}, \bibinfo {author} {\bibfnamefont {A.}~\bibnamefont {Bandyopadhyay}},
  \bibinfo {author} {\bibfnamefont {S.}~\bibnamefont {Datta}}, \bibinfo
  {author} {\bibfnamefont {D.}~\bibnamefont {Bhattacharya}}, \ and\ \bibinfo
  {author} {\bibfnamefont {D.}~\bibnamefont {Jana}},\ }\href {\doibase
  10.1088/1361-648x/ac3075} {\bibfield  {journal} {\bibinfo  {journal} {Journal
  of Physics: Condensed Matter}\ }\textbf {\bibinfo {volume} {34}},\ \bibinfo
  {pages} {053001} (\bibinfo {year} {2021})}\BibitemShut {NoStop}%
\bibitem [{\citenamefont {Benalcazar}\ \emph {et~al.}(2019)\citenamefont
  {Benalcazar}, \citenamefont {Li},\ and\ \citenamefont
  {Hughes}}]{Benalcazar2019}%
  \BibitemOpen
  \bibfield  {author} {\bibinfo {author} {\bibfnamefont {W.~A.}\ \bibnamefont
  {Benalcazar}}, \bibinfo {author} {\bibfnamefont {T.}~\bibnamefont {Li}}, \
  and\ \bibinfo {author} {\bibfnamefont {T.~L.}\ \bibnamefont {Hughes}},\
  }\href {\doibase 10.1103/PhysRevB.99.245151} {\bibfield  {journal} {\bibinfo
  {journal} {Phys. Rev. B}\ }\textbf {\bibinfo {volume} {99}},\ \bibinfo
  {pages} {245151} (\bibinfo {year} {2019})}\BibitemShut {NoStop}%
\bibitem [{\citenamefont {Lopes~dos Santos}\ \emph {et~al.}(2007)\citenamefont
  {Lopes~dos Santos}, \citenamefont {Peres},\ and\ \citenamefont
  {Castro~Neto}}]{PhysRevLett.99.256802}%
  \BibitemOpen
  \bibfield  {author} {\bibinfo {author} {\bibfnamefont {J.~M.~B.}\
  \bibnamefont {Lopes~dos Santos}}, \bibinfo {author} {\bibfnamefont
  {N.~M.~R.}\ \bibnamefont {Peres}}, \ and\ \bibinfo {author} {\bibfnamefont
  {A.~H.}\ \bibnamefont {Castro~Neto}},\ }\href {\doibase
  10.1103/PhysRevLett.99.256802} {\bibfield  {journal} {\bibinfo  {journal}
  {Phys. Rev. Lett.}\ }\textbf {\bibinfo {volume} {99}},\ \bibinfo {pages}
  {256802} (\bibinfo {year} {2007})}\BibitemShut {NoStop}%
\bibitem [{\citenamefont {Leenaerts}\ \emph {et~al.}(2013)\citenamefont
  {Leenaerts}, \citenamefont {Partoens},\ and\ \citenamefont
  {Peeters}}]{Leenaerts2013}%
  \BibitemOpen
  \bibfield  {author} {\bibinfo {author} {\bibfnamefont {O.}~\bibnamefont
  {Leenaerts}}, \bibinfo {author} {\bibfnamefont {B.}~\bibnamefont {Partoens}},
  \ and\ \bibinfo {author} {\bibfnamefont {F.~M.}\ \bibnamefont {Peeters}},\
  }\href {\doibase https://doi.org/10.1063/1.4812977} {\bibfield  {journal}
  {\bibinfo  {journal} {Applied Physics Letters}\ }\textbf {\bibinfo {volume}
  {103}},\ \bibinfo {pages} {013105} (\bibinfo {year} {2013})}\BibitemShut
  {NoStop}%
\bibitem [{\citenamefont {Slater}\ and\ \citenamefont
  {Koster}(1954)}]{Slater1954}%
  \BibitemOpen
  \bibfield  {author} {\bibinfo {author} {\bibfnamefont {J.~C.}\ \bibnamefont
  {Slater}}\ and\ \bibinfo {author} {\bibfnamefont {G.~F.}\ \bibnamefont
  {Koster}},\ }\href {\doibase 10.1103/PhysRev.94.1498} {\bibfield  {journal}
  {\bibinfo  {journal} {Phys. Rev.}\ }\textbf {\bibinfo {volume} {94}},\
  \bibinfo {pages} {1498} (\bibinfo {year} {1954})}\BibitemShut {NoStop}%
\bibitem [{\citenamefont {Chen}\ \emph {et~al.}(2022)\citenamefont {Chen},
  \citenamefont {Zeng}, \citenamefont {Chen}, \citenamefont {Zhao},
  \citenamefont {Sheng},\ and\ \citenamefont {Yang}}]{Chen2022}%
  \BibitemOpen
  \bibfield  {author} {\bibinfo {author} {\bibfnamefont {C.}~\bibnamefont
  {Chen}}, \bibinfo {author} {\bibfnamefont {X.-T.}\ \bibnamefont {Zeng}},
  \bibinfo {author} {\bibfnamefont {Z.}~\bibnamefont {Chen}}, \bibinfo {author}
  {\bibfnamefont {Y.~X.}\ \bibnamefont {Zhao}}, \bibinfo {author}
  {\bibfnamefont {X.-L.}\ \bibnamefont {Sheng}}, \ and\ \bibinfo {author}
  {\bibfnamefont {S.~A.}\ \bibnamefont {Yang}},\ }\href {\doibase
  10.1103/PhysRevLett.128.026405} {\bibfield  {journal} {\bibinfo  {journal}
  {Phys. Rev. Lett.}\ }\textbf {\bibinfo {volume} {128}},\ \bibinfo {pages}
  {026405} (\bibinfo {year} {2022})}\BibitemShut {NoStop}%
\bibitem [{\citenamefont {Ahn}\ \emph {et~al.}(2018)\citenamefont {Ahn},
  \citenamefont {Kim}, \citenamefont {Kim},\ and\ \citenamefont
  {Yang}}]{Ahn2018}%
  \BibitemOpen
  \bibfield  {author} {\bibinfo {author} {\bibfnamefont {J.}~\bibnamefont
  {Ahn}}, \bibinfo {author} {\bibfnamefont {D.}~\bibnamefont {Kim}}, \bibinfo
  {author} {\bibfnamefont {Y.}~\bibnamefont {Kim}}, \ and\ \bibinfo {author}
  {\bibfnamefont {B.-J.}\ \bibnamefont {Yang}},\ }\href {\doibase
  10.1103/PhysRevLett.121.106403} {\bibfield  {journal} {\bibinfo  {journal}
  {Phys. Rev. Lett.}\ }\textbf {\bibinfo {volume} {121}},\ \bibinfo {pages}
  {106403} (\bibinfo {year} {2018})}\BibitemShut {NoStop}%
\bibitem [{\citenamefont {Zou}\ \emph {et~al.}(2018)\citenamefont {Zou},
  \citenamefont {Po}, \citenamefont {Vishwanath},\ and\ \citenamefont
  {Senthil}}]{Zou2018}%
  \BibitemOpen
  \bibfield  {author} {\bibinfo {author} {\bibfnamefont {L.}~\bibnamefont
  {Zou}}, \bibinfo {author} {\bibfnamefont {H.~C.}\ \bibnamefont {Po}},
  \bibinfo {author} {\bibfnamefont {A.}~\bibnamefont {Vishwanath}}, \ and\
  \bibinfo {author} {\bibfnamefont {T.}~\bibnamefont {Senthil}},\ }\href
  {\doibase 10.1103/PhysRevB.98.085435} {\bibfield  {journal} {\bibinfo
  {journal} {Phys. Rev. B}\ }\textbf {\bibinfo {volume} {98}},\ \bibinfo
  {pages} {085435} (\bibinfo {year} {2018})}\BibitemShut {NoStop}%
\bibitem [{\citenamefont {Imhof}\ \emph {et~al.}(2018)\citenamefont {Imhof},
  \citenamefont {Berger}, \citenamefont {Bayer}, \citenamefont {Brehm},
  \citenamefont {Molenkamp}, \citenamefont {Kiessling}, \citenamefont
  {Schindler}, \citenamefont {Lee}, \citenamefont {Greiter}, \citenamefont
  {Neupert},\ and\ \citenamefont {Thomale}}]{circuit2018}%
  \BibitemOpen
  \bibfield  {author} {\bibinfo {author} {\bibfnamefont {S.}~\bibnamefont
  {Imhof}}, \bibinfo {author} {\bibfnamefont {C.}~\bibnamefont {Berger}},
  \bibinfo {author} {\bibfnamefont {F.}~\bibnamefont {Bayer}}, \bibinfo
  {author} {\bibfnamefont {J.}~\bibnamefont {Brehm}}, \bibinfo {author}
  {\bibfnamefont {L.~W.}\ \bibnamefont {Molenkamp}}, \bibinfo {author}
  {\bibfnamefont {T.}~\bibnamefont {Kiessling}}, \bibinfo {author}
  {\bibfnamefont {F.}~\bibnamefont {Schindler}}, \bibinfo {author}
  {\bibfnamefont {C.~H.}\ \bibnamefont {Lee}}, \bibinfo {author} {\bibfnamefont
  {M.}~\bibnamefont {Greiter}}, \bibinfo {author} {\bibfnamefont
  {T.}~\bibnamefont {Neupert}}, \ and\ \bibinfo {author} {\bibfnamefont
  {R.}~\bibnamefont {Thomale}},\ }\href {\doibase 10.1038/s41567-018-0246-1}
  {\bibfield  {journal} {\bibinfo  {journal} {Nature Physics}\ }\textbf
  {\bibinfo {volume} {14}},\ \bibinfo {pages} {925} (\bibinfo {year}
  {2018})}\BibitemShut {NoStop}%
\bibitem [{\citenamefont {Ma}\ \emph {et~al.}(2019)\citenamefont {Ma},
  \citenamefont {Xiao},\ and\ \citenamefont {Chan}}]{Ma2019}%
  \BibitemOpen
  \bibfield  {author} {\bibinfo {author} {\bibfnamefont {G.}~\bibnamefont
  {Ma}}, \bibinfo {author} {\bibfnamefont {M.}~\bibnamefont {Xiao}}, \ and\
  \bibinfo {author} {\bibfnamefont {C.~T.}\ \bibnamefont {Chan}},\ }\href
  {\doibase 10.1038/s42254-019-0030-x} {\bibfield  {journal} {\bibinfo
  {journal} {Nature Reviews Physics}\ }\textbf {\bibinfo {volume} {1}},\
  \bibinfo {pages} {281} (\bibinfo {year} {2019})}\BibitemShut {NoStop}%
\bibitem [{\citenamefont {Xue}\ \emph {et~al.}(2020)\citenamefont {Xue},
  \citenamefont {Ge}, \citenamefont {Sun}, \citenamefont {Wang}, \citenamefont
  {Jia}, \citenamefont {Guan}, \citenamefont {Yuan}, \citenamefont {Chong},\
  and\ \citenamefont {Zhang}}]{Xue2020}%
  \BibitemOpen
  \bibfield  {author} {\bibinfo {author} {\bibfnamefont {H.}~\bibnamefont
  {Xue}}, \bibinfo {author} {\bibfnamefont {Y.}~\bibnamefont {Ge}}, \bibinfo
  {author} {\bibfnamefont {H.-X.}\ \bibnamefont {Sun}}, \bibinfo {author}
  {\bibfnamefont {Q.}~\bibnamefont {Wang}}, \bibinfo {author} {\bibfnamefont
  {D.}~\bibnamefont {Jia}}, \bibinfo {author} {\bibfnamefont {Y.-J.}\
  \bibnamefont {Guan}}, \bibinfo {author} {\bibfnamefont {S.-Q.}\ \bibnamefont
  {Yuan}}, \bibinfo {author} {\bibfnamefont {Y.}~\bibnamefont {Chong}}, \ and\
  \bibinfo {author} {\bibfnamefont {B.}~\bibnamefont {Zhang}},\ }\href
  {\doibase 10.1038/s41467-020-16350-1} {\bibfield  {journal} {\bibinfo
  {journal} {Nature Communications}\ }\textbf {\bibinfo {volume} {11}},\
  \bibinfo {pages} {2442} (\bibinfo {year} {2020})}\BibitemShut {NoStop}%
\bibitem [{\citenamefont {Xue}\ \emph {et~al.}(2019)\citenamefont {Xue},
  \citenamefont {Yang}, \citenamefont {Gao}, \citenamefont {Chong},\ and\
  \citenamefont {Zhang}}]{Xue2019to}%
  \BibitemOpen
  \bibfield  {author} {\bibinfo {author} {\bibfnamefont {H.}~\bibnamefont
  {Xue}}, \bibinfo {author} {\bibfnamefont {Y.}~\bibnamefont {Yang}}, \bibinfo
  {author} {\bibfnamefont {F.}~\bibnamefont {Gao}}, \bibinfo {author}
  {\bibfnamefont {Y.}~\bibnamefont {Chong}}, \ and\ \bibinfo {author}
  {\bibfnamefont {B.}~\bibnamefont {Zhang}},\ }\href
  {https://www.nature.com/articles/s41563-018-0251-x} {\bibfield  {journal}
  {\bibinfo  {journal} {Nature materials}\ }\textbf {\bibinfo {volume} {18}},\
  \bibinfo {pages} {108} (\bibinfo {year} {2019})}\BibitemShut {NoStop}%
\bibitem [{\citenamefont {Ozawa}\ \emph {et~al.}(2019)\citenamefont {Ozawa},
  \citenamefont {Price}, \citenamefont {Amo}, \citenamefont {Goldman},
  \citenamefont {Hafezi}, \citenamefont {Lu}, \citenamefont {Rechtsman},
  \citenamefont {Schuster}, \citenamefont {Simon}, \citenamefont {Zilberberg},\
  and\ \citenamefont {Carusotto}}]{Ozawa2019}%
  \BibitemOpen
  \bibfield  {author} {\bibinfo {author} {\bibfnamefont {T.}~\bibnamefont
  {Ozawa}}, \bibinfo {author} {\bibfnamefont {H.~M.}\ \bibnamefont {Price}},
  \bibinfo {author} {\bibfnamefont {A.}~\bibnamefont {Amo}}, \bibinfo {author}
  {\bibfnamefont {N.}~\bibnamefont {Goldman}}, \bibinfo {author} {\bibfnamefont
  {M.}~\bibnamefont {Hafezi}}, \bibinfo {author} {\bibfnamefont
  {L.}~\bibnamefont {Lu}}, \bibinfo {author} {\bibfnamefont {M.~C.}\
  \bibnamefont {Rechtsman}}, \bibinfo {author} {\bibfnamefont {D.}~\bibnamefont
  {Schuster}}, \bibinfo {author} {\bibfnamefont {J.}~\bibnamefont {Simon}},
  \bibinfo {author} {\bibfnamefont {O.}~\bibnamefont {Zilberberg}}, \ and\
  \bibinfo {author} {\bibfnamefont {I.}~\bibnamefont {Carusotto}},\ }\href
  {\doibase 10.1103/RevModPhys.91.015006} {\bibfield  {journal} {\bibinfo
  {journal} {Rev. Mod. Phys.}\ }\textbf {\bibinfo {volume} {91}},\ \bibinfo
  {pages} {015006} (\bibinfo {year} {2019})}\BibitemShut {NoStop}%
\bibitem [{\citenamefont {Mittal}\ \emph {et~al.}(2019)\citenamefont {Mittal},
  \citenamefont {Orre}, \citenamefont {Zhu}, \citenamefont {Gorlach},
  \citenamefont {Poddubny},\ and\ \citenamefont {Hafezi}}]{Mittal2019}%
  \BibitemOpen
  \bibfield  {author} {\bibinfo {author} {\bibfnamefont {S.}~\bibnamefont
  {Mittal}}, \bibinfo {author} {\bibfnamefont {V.~V.}\ \bibnamefont {Orre}},
  \bibinfo {author} {\bibfnamefont {G.}~\bibnamefont {Zhu}}, \bibinfo {author}
  {\bibfnamefont {M.~A.}\ \bibnamefont {Gorlach}}, \bibinfo {author}
  {\bibfnamefont {A.}~\bibnamefont {Poddubny}}, \ and\ \bibinfo {author}
  {\bibfnamefont {M.}~\bibnamefont {Hafezi}},\ }\href {\doibase
  10.1038/s41566-019-0452-0} {\bibfield  {journal} {\bibinfo  {journal} {Nature
  Photonics}\ }\textbf {\bibinfo {volume} {13}},\ \bibinfo {pages} {692}
  (\bibinfo {year} {2019})}\BibitemShut {NoStop}%
\bibitem [{\citenamefont {Prodan}\ and\ \citenamefont
  {Prodan}(2009)}]{mechanical2009}%
  \BibitemOpen
  \bibfield  {author} {\bibinfo {author} {\bibfnamefont {E.}~\bibnamefont
  {Prodan}}\ and\ \bibinfo {author} {\bibfnamefont {C.}~\bibnamefont
  {Prodan}},\ }\href {\doibase 10.1103/PhysRevLett.103.248101} {\bibfield
  {journal} {\bibinfo  {journal} {Phys. Rev. Lett.}\ }\textbf {\bibinfo
  {volume} {103}},\ \bibinfo {pages} {248101} (\bibinfo {year}
  {2009})}\BibitemShut {NoStop}%
\bibitem [{\citenamefont {Dalibard}\ \emph {et~al.}(2011)\citenamefont
  {Dalibard}, \citenamefont {Gerbier}, \citenamefont {Juzeli\ifmmode~\else
  \={u}\fi{}nas},\ and\ \citenamefont {\"Ohberg}}]{ColdAtomRMP}%
  \BibitemOpen
  \bibfield  {author} {\bibinfo {author} {\bibfnamefont {J.}~\bibnamefont
  {Dalibard}}, \bibinfo {author} {\bibfnamefont {F.}~\bibnamefont {Gerbier}},
  \bibinfo {author} {\bibfnamefont {G.}~\bibnamefont {Juzeli\ifmmode~\else
  \={u}\fi{}nas}}, \ and\ \bibinfo {author} {\bibfnamefont {P.}~\bibnamefont
  {\"Ohberg}},\ }\href {\doibase 10.1103/RevModPhys.83.1523} {\bibfield
  {journal} {\bibinfo  {journal} {Rev. Mod. Phys.}\ }\textbf {\bibinfo {volume}
  {83}},\ \bibinfo {pages} {1523} (\bibinfo {year} {2011})}\BibitemShut
  {NoStop}%
\bibitem [{\citenamefont {Kane}\ and\ \citenamefont
  {Mele}(2005)}]{Kane2005Graphene}%
  \BibitemOpen
  \bibfield  {author} {\bibinfo {author} {\bibfnamefont {C.~L.}\ \bibnamefont
  {Kane}}\ and\ \bibinfo {author} {\bibfnamefont {E.~J.}\ \bibnamefont
  {Mele}},\ }\href {\doibase 10.1103/PhysRevLett.95.226801} {\bibfield
  {journal} {\bibinfo  {journal} {Phys. Rev. Lett.}\ }\textbf {\bibinfo
  {volume} {95}},\ \bibinfo {pages} {1} (\bibinfo {year} {2005})}\BibitemShut
  {NoStop}%
\bibitem [{\citenamefont {Kresse}\ and\ \citenamefont
  {Hafner}(1994)}]{Kresse1994}%
  \BibitemOpen
  \bibfield  {author} {\bibinfo {author} {\bibfnamefont {G.}~\bibnamefont
  {Kresse}}\ and\ \bibinfo {author} {\bibfnamefont {J.}~\bibnamefont
  {Hafner}},\ }\href {\doibase 10.1103/PhysRevB.49.14251} {\bibfield  {journal}
  {\bibinfo  {journal} {Phys. Rev. B}\ }\textbf {\bibinfo {volume} {49}},\
  \bibinfo {pages} {14251} (\bibinfo {year} {1994})}\BibitemShut {NoStop}%
\bibitem [{\citenamefont {Kresse}\ and\ \citenamefont
  {Furthm\"uller}(1996)}]{Kresse1996}%
  \BibitemOpen
  \bibfield  {author} {\bibinfo {author} {\bibfnamefont {G.}~\bibnamefont
  {Kresse}}\ and\ \bibinfo {author} {\bibfnamefont {J.}~\bibnamefont
  {Furthm\"uller}},\ }\href {\doibase 10.1103/PhysRevB.54.11169} {\bibfield
  {journal} {\bibinfo  {journal} {Phys. Rev. B}\ }\textbf {\bibinfo {volume}
  {54}},\ \bibinfo {pages} {11169} (\bibinfo {year} {1996})}\BibitemShut
  {NoStop}%
\bibitem [{\citenamefont {Bl\"ochl}(1994)}]{PAW}%
  \BibitemOpen
  \bibfield  {author} {\bibinfo {author} {\bibfnamefont {P.~E.}\ \bibnamefont
  {Bl\"ochl}},\ }\href {\doibase 10.1103/PhysRevB.50.17953} {\bibfield
  {journal} {\bibinfo  {journal} {Phys. Rev. B}\ }\textbf {\bibinfo {volume}
  {50}},\ \bibinfo {pages} {17953} (\bibinfo {year} {1994})}\BibitemShut
  {NoStop}%
\bibitem [{\citenamefont {Perdew}\ \emph {et~al.}(1996)\citenamefont {Perdew},
  \citenamefont {Burke},\ and\ \citenamefont {Ernzerhof}}]{PBE}%
  \BibitemOpen
  \bibfield  {author} {\bibinfo {author} {\bibfnamefont {J.~P.}\ \bibnamefont
  {Perdew}}, \bibinfo {author} {\bibfnamefont {K.}~\bibnamefont {Burke}}, \
  and\ \bibinfo {author} {\bibfnamefont {M.}~\bibnamefont {Ernzerhof}},\ }\href
  {\doibase 10.1103/PhysRevLett.77.3865} {\bibfield  {journal} {\bibinfo
  {journal} {Phys. Rev. Lett.}\ }\textbf {\bibinfo {volume} {77}},\ \bibinfo
  {pages} {3865} (\bibinfo {year} {1996})}\BibitemShut {NoStop}%
\bibitem [{\citenamefont {Grimme}\ \emph {et~al.}(2010)\citenamefont {Grimme},
  \citenamefont {Antony}, \citenamefont {Ehrlich},\ and\ \citenamefont
  {Krieg}}]{DFTD3}%
  \BibitemOpen
  \bibfield  {author} {\bibinfo {author} {\bibfnamefont {S.}~\bibnamefont
  {Grimme}}, \bibinfo {author} {\bibfnamefont {J.}~\bibnamefont {Antony}},
  \bibinfo {author} {\bibfnamefont {S.}~\bibnamefont {Ehrlich}}, \ and\
  \bibinfo {author} {\bibfnamefont {H.}~\bibnamefont {Krieg}},\ }\href
  {\doibase https://doi.org/10.1063/1.3382344} {\bibfield  {journal} {\bibinfo
  {journal} {The Journal of Chemical Physics}\ }\textbf {\bibinfo {volume}
  {132}},\ \bibinfo {pages} {154104} (\bibinfo {year} {2010})}\BibitemShut
  {NoStop}%
\bibitem [{\citenamefont {Grimme}\ \emph {et~al.}(2011)\citenamefont {Grimme},
  \citenamefont {Ehrlich},\ and\ \citenamefont {Goerigk}}]{DFTD3BJ}%
  \BibitemOpen
  \bibfield  {author} {\bibinfo {author} {\bibfnamefont {S.}~\bibnamefont
  {Grimme}}, \bibinfo {author} {\bibfnamefont {S.}~\bibnamefont {Ehrlich}}, \
  and\ \bibinfo {author} {\bibfnamefont {L.}~\bibnamefont {Goerigk}},\ }\href
  {\doibase https://doi.org/10.1002/jcc.21759} {\bibfield  {journal} {\bibinfo
  {journal} {Journal of Computational Chemistry}\ }\textbf {\bibinfo {volume}
  {32}},\ \bibinfo {pages} {1456} (\bibinfo {year} {2011})}\BibitemShut
  {NoStop}%
\bibitem [{\citenamefont {Moon}\ and\ \citenamefont
  {Koshino}(2013)}]{Moon2013}%
  \BibitemOpen
  \bibfield  {author} {\bibinfo {author} {\bibfnamefont {P.}~\bibnamefont
  {Moon}}\ and\ \bibinfo {author} {\bibfnamefont {M.}~\bibnamefont {Koshino}},\
  }\href {\doibase 10.1103/PhysRevB.87.205404} {\bibfield  {journal} {\bibinfo
  {journal} {Phys. Rev. B}\ }\textbf {\bibinfo {volume} {87}},\ \bibinfo
  {pages} {205404} (\bibinfo {year} {2013})}\BibitemShut {NoStop}%
\bibitem [{\citenamefont {Wang}\ and\ \citenamefont {Moore}(2019)}]{Moore2019}%
  \BibitemOpen
  \bibfield  {author} {\bibinfo {author} {\bibfnamefont {Y.-Q.}\ \bibnamefont
  {Wang}}\ and\ \bibinfo {author} {\bibfnamefont {J.~E.}\ \bibnamefont
  {Moore}},\ }\href {\doibase 10.1103/PhysRevB.99.155102} {\bibfield  {journal}
  {\bibinfo  {journal} {Phys. Rev. B}\ }\textbf {\bibinfo {volume} {99}},\
  \bibinfo {pages} {155102} (\bibinfo {year} {2019})}\BibitemShut {NoStop}%
\end{thebibliography}%


\end{document}